\documentclass[aps, prd, twocolumn, notitlepage, nofootinbib,showpacs]{revtex4-1}
\usepackage{graphicx}
\usepackage{float}
\usepackage{amsmath}
\usepackage{color}
\usepackage{tensor}
\usepackage{epstopdf}
\usepackage{xcolor}
\usepackage{wasysym}
\usepackage{hyperref}

\hypersetup{
    colorlinks,
    linkcolor={blue!50!black},
    citecolor={red!50!black},
    urlcolor={blue!80!black}
}

\usepackage{multibib}

\newcommand{\be}{\begin{equation}}
\newcommand{\ee}{\end{equation}}
\newcommand{\ba}{\begin{eqnarray}}
\newcommand{\ea}{\end{eqnarray}}

\newcommand{\beq}{\begin{equation}}
\newcommand{\eeq}{\end{equation}}
\newcommand{\beqa}{\begin{eqnarray}}
\newcommand{\eeqa}{\end{eqnarray}}
\newcommand{\nn}{\nonumber}
\graphicspath{{plots/}}

\begin{document}

\author{Amjad Ashoorioon}
\email{amjad@ipm.ir}
\affiliation{School of Physics, Institute for Research in Fundamental Sciences (IPM), P.O. Box 19395-5531, Tehran, Iran.}

\author{Mohammad Bagher Jahani Poshteh}
\email{jahani@ipm.ir}
\affiliation{School of Physics, Institute for Research in Fundamental Sciences (IPM), P.O. Box 19395-5531, Tehran, Iran.}

\author{Orlando Luongo}
\email{orlando.luongo@unicam.it}
\affiliation{Universit\`a di Camerino, Divisione di Fisica, Via Madonna delle carceri 9, 62032 Camerino, Italy.}
\affiliation{SUNY Polytechnic Institute, 13502 Utica, New York, USA.}
\affiliation{INFN, Sezione di Perugia, Perugia, 06123, Italy.}
\affiliation{INAF - Osservatorio Astronomico di Brera, Milano, Italy.}
\affiliation{Al-Farabi Kazakh National University, Al-Farabi av. 71, 050040 Almaty, Kazakhstan.}

\title{Particle motion and accretion disk around rotating accelerating black holes}

\begin{abstract}
We investigate the motion of test particles and the properties of accretion disks around rotating, accelerating black hole spacetimes with non-zero cosmological constants. To do so, we explore the impact of both rotation and acceleration on massive particle dynamics in motion around these configurations. In this respect, we derive the geodesic equations in the Pleba{\'n}ski-Demia{\'n}ski spacetime and analyse the characteristics of circular orbits, including the innermost and outermost stable circular orbits. Accordingly, we find that the acceleration parameter significantly influences the radii and angular momenta of these orbits. Further, we examine the precession of non-circular orbits and find that the precession is in the same direction as the black hole rotation regardless of the direction of the black hole spin. We also investigate the thermal spectra of geometrically thin, optically thick accretion disks described by the Novikov-Thorne model. We consider both co-rotating and counter-rotating disks and show that the radiative flux and luminosity are significantly larger for co-rotating disks. We compare the spectral features and luminosity of accretion disks around accelerating black holes with those around non-accelerating Kerr black holes, revealing that acceleration generally reduces the luminosity and flux of emitted light. Our results provide insights into the complex interplay between black hole acceleration, rotation, and the surrounding accretion disk dynamics.
\end{abstract}


\maketitle
\tableofcontents

\section{Introduction}

The motion of test particles in black hole spacetime will come out very handy to investigate the properties and effects of such extreme gravitational objects and test/verify various theories of gravity. By solving the equations of motion of test particles in the different metrics of black holes, we can get information about the mass, charge, spin, event horizon, ergosphere, and singularity of a black hole. We will also be comparing the predictions of different theories of gravitation, including the so called modified or extended theories of gravity, quantum gravity, and so on, in search of possible departures from general relativity.

Additionally, the motion of test particles finds practical applications in astrophysics and cosmology, such as detecting gravitational waves from binary black hole mergers using the post-Newtonian approximation \cite{Damour:2002vi}, modeling accretion disks and jets around black holes using magnetohydrodynamics \cite{ustyugova2000poynting}, and simulating formation and evolution of black holes in various environments within numerical relativity \cite{Shibata:1999zs}. Motion of test particles on timelike geodesics around black holes has been widely studied in literature
\footnote{For example, motion of particles in spherically symmetric spacetime in de Sitter (dS) and anti de Sitter (AdS) backgrounds has been studied, respectively, in \cite{jaklitsch1989} and \cite{Cruz:2004ts}. The problem has also been discussed for Reissner-Nordstr\"om (A)dS black holes in \cite{Stuchlik:2002tj} (and with more details in \cite{Pugliese:2011py} in flat background). Timelike geodesics have also been investigated for Kerr \cite{Fayos:2007ks}, Kerr-dS \cite{Stuchlik:2003dt}, and Kerr-Newman-(A)dS \cite{Soroushfar:2016esy} spacetimes. In \cite{Das:2020yxw} the authors study timelike geodesics around charged rotating black holes in the presence of perfect fluid dark matter and in \cite{Wilkins:1972rs,Teo:2020sey} spherical orbits around Kerr black holes have been explored. Further, dynamical properties related to repulsive gravitational effects can be highlighted through the motion of test particles \cite{Luongo:2014qoa,Luongo:2015zaa,Luongo:2023aib}.}.

For the case of extremely high-spin Kerr black holes the authors of \cite{Compere:2020eat} have studied the geodesics in the near horizon region and provide a classification of particle motion. They also have found exact expressions for the innermost stable and bound spherical orbits, and show how they are related by a conformal transformation and a discrete symmetry. The paper \cite{Kapec:2019hro} studies the dynamics of test particles in the vicinity of a rotating black hole with maximal or near-maximal spin.

A remarkable case is provided by \emph{accelerating black holes}. These represent solutions of the Einstein field equations that describe black holes that are moving with a constant acceleration due to a conical deficit along their polar axis \cite{kinnersley1970}. This conical deficit could be caused by a cosmic string, a hypothetical one-dimensional defect in spacetime that have tension and energy \cite{vilenkin1985,hr,emparan1995,eardley1995,Ashoorioon:2014ipa,ashoorioon2021}. If it passes through a black hole, the cosmic string can cause an acceleration of the black hole. Black holes can be accelerating because of the existence of an external magnetic field or a positive cosmological constant too \cite{ashoorioon2021}. Black hole acceleration will have a big part to do with changing their numerous aspects, such as their gravitational lensing \cite{Ashoorioon:2022zgu,Ashoorioon:2021gjs}.

In this scenario, the physics of accretion disks around black holes can mainly be in support of the motion of circular orbits and disclose the main characteristics of a given black hole configuration. The importance of accretion disks that form around black holes lies in a number of aspects. First, they are the main sources of electromagnetic radiation from black hole systems, meaning that we can observe and study them in most cases. Accretion disks could radiate over a very large range in frequency --- basically from radio waves up to X-rays --- depending on the mass and temperature of the disk and the black hole \cite{Narayan:1993bd}. It is expected that such radiation could carry information about the black hole properties, such as mass, spin, and charge, and the surroundings of the black hole. It can also reveal relativistic effects like gravitational redshift, Doppler shift, and gravitational lensing \cite{Tanaka:1995en}. Second, accretion disks around black holes are important in understanding the physics of matter and gravity under extreme conditions. Matter in an accretion disk is subjected to very strong gravitational, thermal, magnetic, and radiative forces, which produce a number of highly nonlinear complex phenomena. This includes turbulence, viscosity, magnetohydrodynamic instabilities, shocks, jets, outflows, and winds\footnote{Such physics of the accretion disks applies to neutron stars, white dwarfs, protoplanetary disks, and active galactic nuclei. In addition, the accretion disks around black holes open up the possibility for understanding the origin and evolution of the black holes themselves by detailing how they gain mass through the accreting of matter, how they interact with their environments, and how they have a role in the formation and structuring of galaxies.}  \cite{Kato:2022cur,Kadowaki:2018dtb,Andreoni:2022afu,Kosec:2023qva}.

In this respect, throughout recent years several papers have reviewed the theoretical and observational aspects of accretion disks around black holes, e.g. Abramowicz and Fragile discussed the basic concepts of black hole accretion disk theory, the different models of accretion disks, and the issues of stability, oscillations, and jets \cite{Abramowicz:2011xu}. They also presented some applications of the theory to measure black hole mass and spin, to study black hole spectral states, and to explain quasi-periodic oscillations.

Accretion disks around naked singularities have been studied in \cite{Joshi:2013dva,Shaikh:2019hbm}. The effects of cosmological constant on accretion disks are investigated in \cite{Stuchlik:2005euw}. The luminosity of accretion disks in the presence of dark matter has been considered in \cite{Boshkayev:2020kle,Kurmanov:2021uqv,Boshkayev:2022vlv}. In \cite{DAgostino:2022ckg} the accretion disk luminosity has been used to constrain primordial black holes as a fraction of dark matter. More recently, the accretion disk in rotating regular black holes \cite{Boshkayev:2023fft}, Hartle-Thorne spacetime \cite{Boshkayev:2023ipb}, and compact objects with quadrupole \cite{Boshkayev:2021chc} has been worked out.

Motivated by all the aforementioned scenarios, we here investigate the motion of test particles around \emph{rotating accelerating black holes} in a background with non-zero cosmological constant. More generally, we extend the work in which  timelike geodesics in non-rotating accelerating black hole spacetime have been explored \cite{JahaniPoshteh:2022yei}. There, investigation of circular orbits has shown that, along with innermost stable circular orbit (ISCO), there also exists an outermost stable circular orbit (OSCO) in flat, dS, and AdS backgrounds. It has also been shown that the precession of perihelion would increase by increasing the acceleration. Hence, in this paper we study the joint effect of both  rotation and acceleration on the motion of massive test particles in a spacetime with non-zero cosmological constant. In particular we consider the black hole at the Galactic center, Sgr A*, and study the corresponding motion of S2 star around it. Sgr A*, as well as other supermassive black holes, might be accelerating
\footnote{Remarkably, it has been speculated that Sgr A* could be connected to a cosmic string, see e.g. \cite{morris2017}.} \cite{vilenkin2018}. Nevertheless, in singling out the kind of acceleration, structure formation plays a significant role. Precisely, the underlying acceleration should be small enough to enable it to be captured by the Galaxy \cite{JahaniPoshteh:2022yei,vilenkin2018}. We therefore focus on {\it slowly}-accelerating rotating black holes.

Afterwards, we also study accretion disk around rotating slowly-accelerating black holes in different backgrounds with positive, negative, or zero cosmological constant. Precisely, we consider geometrically thin and optically thick accretion disks which are described by Novikov-Thorne model \cite{Novikov:1973,Page:1974he,Thorne:1974ve}. In this model, essentially based on general relativity, the disk is assumed to be in the equatorial plane of the black hole and consists of test particles moving in circular orbits around the black hole itself. Accordingly, we explore the spectral features of accretion disk around rotating accelerating black holes and compare them with corresponding quantities of accretion disk around non-accelerating Kerr black hole. Finally, we focus on how and whether the  black hole acceleration may increase or decrease the net luminosity and flux of light from accretion disk.

The paper is organized as follows. In the next section we present the derivation of the equations governing the motion of massive particles around accelerating Kerr black holes in the presence of a cosmological constant. In Sec.~\ref{sec:circular}, we examine the circular orbits in the rotating accelerating black hole spacetime. In Sec.~\ref{sec:precession}, we study non-circular motion of test particles around rotating accelerating black holes and analyze the precession of orbits for both co-rotating and counter-rotating cases. In Sec.~\ref{sec:accretion}, we investigate the thermal spectrum of the accretion disk, and conclude our paper in Sec.~\ref{sec:con}.

\section{Geodesic equations in Pleba{\'n}ski-Demia{\'n}ski spacetime}\label{sec:geodesic}

Rotating accelerating black holes are described by Pleba{\'n}ski-Demia{\'n}ski metric \cite{plebanski1976}. This metric is a general class of solutions of the Einstein-Maxwell equations with a cosmological constant. It is a type D metric, which means that it has two repeated principal null directions that are also eigenvectors of the Weyl tensor \cite{VandenBergh:2020lvf}.

\subsection{The metric}

In its general form, the Pleba{\'n}ski-Demia{\'n}ski metric has seven parameters: mass, spin, electric charge, magnetic charge, NUT charge, acceleration, and cosmological constant (we set the electric, magnetic, and NUT charges to zero). The metric can be written as \cite{Griffiths:2005qp,Podolsky:2006px}\\

\be\label{eqn:met}
ds^2=g_{tt}dt^2+g_{rr}dr^2+g_{\theta\theta}d\theta^2+g_{\phi\phi}d\phi^2+2g_{t\phi}dtd\phi,
\ee
with
\ba
g_{tt}&=&-\frac{Q-a^2 P \sin ^2(\theta )}{\rho ^2 \Omega ^2},\\
g_{rr}&=&\frac{\rho ^2}{Q \Omega ^2},\\
g_{\theta\theta}&=&\frac{\rho ^2}{P \Omega ^2},\\
g_{\phi\phi}&=&-\frac{\sin ^2(\theta ) \left(a^2 Q \sin ^2(\theta )-P \left(a^2+r^2\right)^2\right)}{\rho ^2 \Omega ^2},\\
g_{t\phi}&=&-\frac{a \sin ^2(\theta ) \left(P \left(a^2+r^2\right)-Q\right)}{\rho ^2 \Omega ^2},
\ea
where
\ba
\Omega&=&1-\alpha  r \cos (\theta ),\\
\rho^2&=&a^2 \cos ^2(\theta )+r^2,\\
P&=&\cos ^2(\theta ) \left(\alpha ^2 a^2+\frac{a^2 \Lambda }{3}\right)-2 \alpha  m \cos (\theta )+1,\\
Q&=&\left(1-\alpha ^2 r^2\right) \left(a^2-2 m r+r^2\right)-\frac{1}{3} \Lambda  r^2 \left(a^2+r^2\right).\nn\\
\ea
$a$ and $\alpha$ are respectively the rotation and acceleration parameters. $\Lambda$ is the cosmological constant which can be positive (for dS background), negative (for AdS background), or zero (for flat background). Also, $\theta$ is the azimuthal angle and $m$ shows the mass of black hole. There is a deficit angle in $\phi$ associated with conical singularities at the poles, and $-C_0\pi \leq \phi \leq C_0\pi$. $C_0$ can be chosen so as to eliminate one of the singularities. By taking $C_0=(1+2\alpha m)^{-1}$, we can eliminate the singularity at the pole $\theta = 0$ \cite{lim2014}. On the other hand, taking $C_0=(1-2\alpha m)^{-1}$ would eliminate the singularity at the pole $\theta = \pi$. We take the former possibility for $C_0$.

\begin{figure}[htp]
	\centering
	\includegraphics[width=0.45\textwidth]{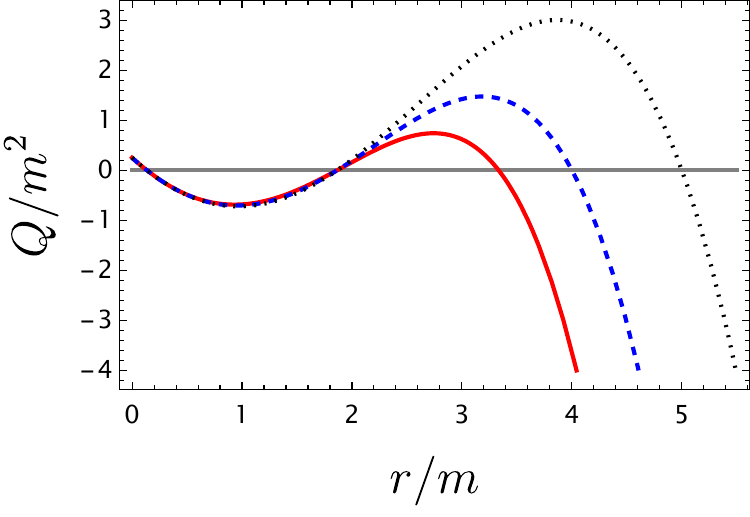}
	\includegraphics[width=0.45\textwidth]{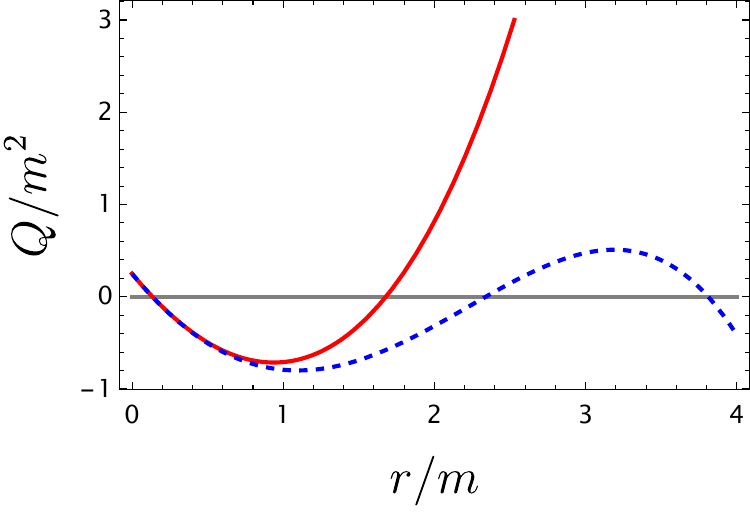}
	\caption{\textit{Top}: Horizons of rotating accelerating black holes in flat background, i.e.~$\Lambda = 0$, which are given by the zeros of the function $Q$. We have $m \alpha = 0.3$ for the solid red plot, $m \alpha = 0.25$ for the dashed blue plot, and $m \alpha = 0.2$ for the dotted black plot. \textit{Bottom}: Horizons of an rotating accelerating black hole in anti de Sitter background with $m^2 \Lambda = -0.1$ (solid red plot) and de Sitter background with $m^2 \Lambda = 0.1$ (dashed blue plot). We have taken $a/m = 0.5$ in both panels.}
	\label{fig:Qplot}
\end{figure}

The horizons of the spacetime can be found from the equation $Q=0$. In the top panel of Fig.~\ref{fig:Qplot} we see that in the flat background there are three horizons in the rotating accelerating black hole spacetime. The smallest root of $Q=0$ indicates the inner horizon, the middle root indicates the event horizon, and the largest root shows the acceleration horizon. The inner and event horizons do not change significantly with the accelerating parameter $\alpha$. However, as $\alpha$ decreases the acceleration horizon would be in a larger radius. In the bottom panel of Fig.~\ref{fig:Qplot} we see that in AdS background with suitable cosmological constant there exist only two horizons. In fact, one can show that for $\Lambda < -3\alpha^2$ the acceleration horizon disappears. In dS background, again we have three horizons. The outermost one can be called a cosmological horizon or acceleration horizon. In the rest of the paper we would be only interested in the region between the event horizon and acceleration horizon. This region is the domain of outer communication for which $Q>0$ --- this implies $g_{rr}>0$.

\subsection{Kinematic quantities}

Here we obtain the equations governing the motion of massive particles around accelerating Kerr black holes in the presence of a cosmological constant. Specifically, we consider the motion of S2 star around Sgr A*. Since the acceleration of the latter is very small, an argument similar to that in \cite{JahaniPoshteh:2022yei} leads us to the conclusion that Sgr A* acceleration should be $\alpha \lesssim 3.29 \times 10^{-26} {\rm m}^{-1}$ and that S2 can be considered as orbiting in the equatorial plane of the Sgr A*.

The geodesic equations in the equatorial plane $\theta=\pi/2$, can be found by using the Lagrangian
\be\label{eqn:lag_geo}
2\mathcal{L} = g_{\mu\nu}\dot{x}^\mu\dot{x}^\nu= g_{tt}\dot{t}^2+g_{rr}\dot{r}^2+g_{\phi\phi}\dot{\phi}^2+2g_{t\phi}\dot{t}\dot{\phi} = -\mu^2,
\ee
where dot shows differentiation with respect to the proper time of test particle with rest mass $\mu > 0$.
The constants of motion can be found by using the Lagrangian \eqref{eqn:lag_geo} as
\ba
E &=& p_t = -\frac{\partial \mathcal{L}}{\partial \dot{t}} = -\left(g_{tt}\dot{t}+g_{t\phi}\dot{\phi}\right), \label{eqn:E}\\
L &=& p_\phi = \frac{\partial \mathcal{L}}{\partial \dot{\phi}} = g_{\phi\phi}\dot{\phi}+g_{t\phi}\dot{t}. \label{eqn:L_z}
\ea
Here $p_\nu$ is the canonical momentum associated to the coordinate $\nu$. $E$ and $L\equiv L_z$ are the energy and (third component of) angular momentum as measured by a locally non-rotating observer. Note that they cannot be always taken as the energy and angular momentum as measured by an observer at infinity since the spacetime can be non-flat.

Solving Eqs.~\eqref{eqn:E} and \eqref{eqn:L_z} for $\dot{t}$ and $\dot{\phi}$, and plugging them into the Lagrangian \eqref{eqn:lag_geo}, we find
\be\label{eqn:grdot}
g_{rr}\dot{r}^2 = \frac{g_{\phi\phi}\tilde{E}^2+2g_{t\phi}\tilde{L}\tilde{E}+g_{tt}\tilde{L}^2}{g_{t\phi}^2-g_{\phi\phi}g_{tt}}-1,
\ee
where $\tilde{E}$ and $\tilde{L}$ are respectively the energy and angular momentum per unit rest mass of the test particle, and dot now denotes differentiation with respect to the proper time per unit rest mass of the particle.

In the domain of outer communication $g_{rr}$ is positive. Therefore, the right hand side of Eq.~\eqref{eqn:grdot} should be positive. The roots of this expression are
\be
\tilde{E}^{(\pm)}=\frac{\left(g_{t\phi}^2- g_{tt} g_{\phi \phi}\right) \left(\pm\frac{ 2 \sqrt{-g_{\phi \phi}-\tilde{L}^2}}{\sqrt{g_{tt} g_{\phi \phi}-g_{t\phi}^2}}-\frac{2 g_{t\phi}\tilde{L}}{g_{t\phi}^2-g_{tt} g_{\phi \phi}}\right)}{2 g_{\phi \phi}}.
\ee
For the right hand side of Eq.~\eqref{eqn:grdot} to be positive, the energy $\tilde{E}$ should satisfy either $\tilde{E}>\tilde{E}^{(+)}$ or $\tilde{E}<\tilde{E}^{(-)}$. In the latter case the energy would be negative and thus this case is ruled out. The relevant effective potential is therefore $\tilde{V}=\tilde{E}^{(+)}$.\\


\section{Circular motion}\label{sec:circular}

\begin{figure}[htp]
	\centering
	\includegraphics[width=0.45\textwidth]{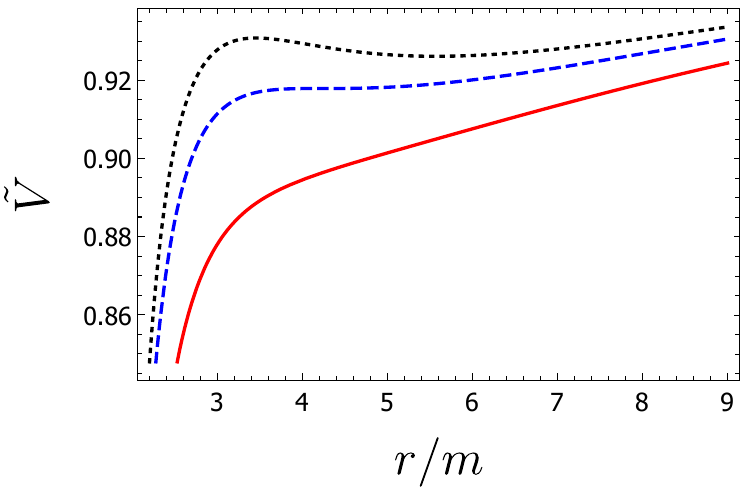}
	\includegraphics[width=0.45\textwidth]{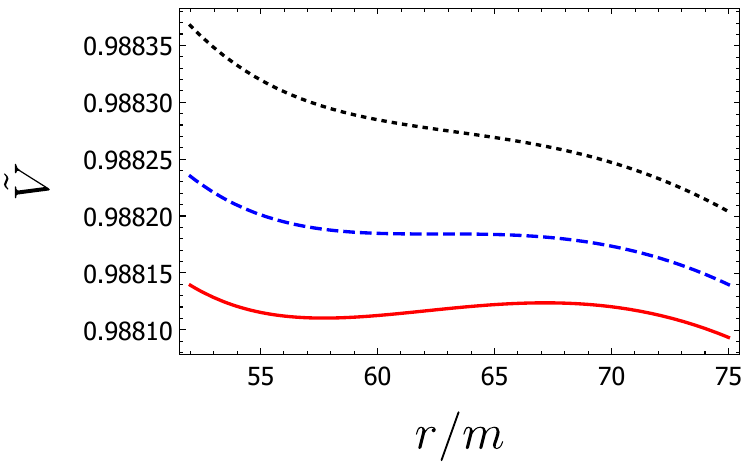}
	\caption{\textit{Top}: The effective potential for co-rotating particles in the region near the black hole horizon with $\tilde{L}/m = 2.7$ for solid red plot, $\tilde{L}/m =\tilde{L}_{\rm ISCO}/m \simeq 2.90276$ for dashed blue plot, and $\tilde{L}/m = 3$ for dotted black plot. The radius of ISCO is $r_{\rm ISCO}/m \simeq 4.23338$. \textit{Bottom}: The effective potential for co-rotating particles in the region far from the black hole with $\tilde{L}/m = 7$ for solid red plot, $\tilde{L}/m = \tilde{L}_{\rm OSCO}/m \simeq 7.03792$ for dashed blue plot, and $\tilde{L}/m = 7.09$ for dotted black plot. The radius of OSCO is $r_{\rm OSCO}/m \simeq 62.5489$. We have taken $a/m = 0.5$, $\theta = \pi/2$, $m\alpha = 10^{-3}$, and $m^2\Lambda = 10^{-52}$ in both panels.}
	\label{fig:pot}
\end{figure}

The extrema of the effective potential $\tilde{V}$ indicate circular orbits. Also, the inflection points of $\tilde{V}$ show last stable circular orbits. In Fig.~\ref{fig:pot}, by assuming $\tilde{L}>0$, we see that for rotating accelerating black hole spacetime there are two inflection points in the plot of the effective potential. This can happen for dS, flat, or AdS background --- for non-accelerating black holes this happens only in dS background \cite{howes1979existence,Boonserm:2019nqq}. One of the inflection points is close to the event horizon (the top panel of Fig.~\ref{fig:pot}) and is associated with ISCO. The other inflection point is in the relatively far distances (the bottom panel of Fig.~\ref{fig:pot}) and shows OSCO. In the plots of this section we take the cosmological constant to be $\Lambda = 10^{-52} {\rm m}^{-2}$, which is approximately the observed value of the cosmological constant \cite{Padmanabhan:2002ji}.

\begin{figure}[htp]
	\centering
	\includegraphics[width=0.45\textwidth]{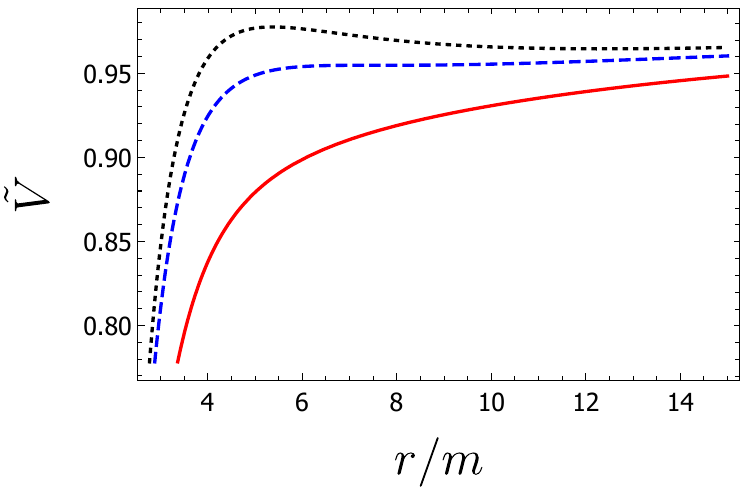}
	\includegraphics[width=0.45\textwidth]{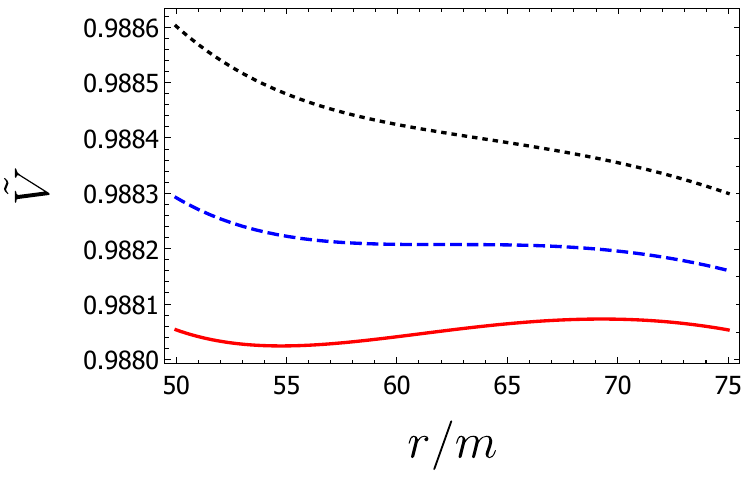}
	\caption{\textit{Top}: The effective potential for counter-rotating particles in the region near the black hole horizon with $\tilde{L}/m = -3 $ for solid red plot, $\tilde{L}/m =\tilde{L}_{\rm ISCO}/m \simeq -3.88349$ for dashed blue plot, and $\tilde{L}/m = -4.2$ for dotted black plot. The radius of ISCO is $r_{\rm ISCO}/m \simeq 7.55923$. \textit{Bottom}: The effective potential for counter-rotating particles in the region far from the black hole with $\tilde{L}/m = -7$ for solid red plot, $\tilde{L}/m = \tilde{L}_{\rm OSCO}/m \simeq -7.08774$ for dashed blue plot, and $\tilde{L}/m = -7.2$ for dotted black plot. The radius of OSCO is $r_{\rm OSCO}/m \simeq 62.2986$. We have taken $a/m = 0.5$, $\theta = \pi/2$, $m\alpha = 10^{-3}$, and $m^2\Lambda = 10^{-52}$ in both panels.}
	\label{fig:retropot}
\end{figure}

For the case in which $\tilde{L}<0$, i.e.~particles counter-rotating with the black hole, we also have two inflection points in the effective potential. This can be observed in Fig.~\ref{fig:retropot}. Again the inflection point near the event horizon shows ISCO and the one far from the black hole shows OSCO. We can find that, compared to co-rotating case, for the counter-rotating particles the radius of ISCO and the magnitude of the angular momentum in this orbit are larger. We also find that the radius of OSCO is smaller than that of co-rotating case and the magnitude of the angular momentum is larger.

\begin{figure}[htp]
	\centering
	\includegraphics[width=0.45\textwidth]{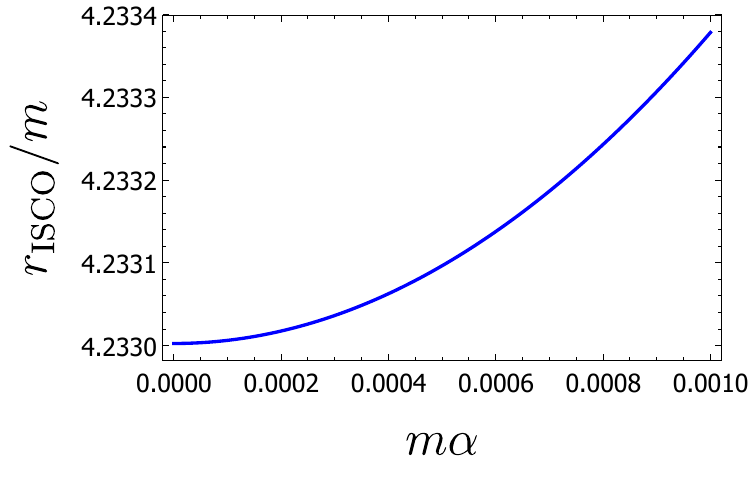}
	\includegraphics[width=0.45\textwidth]{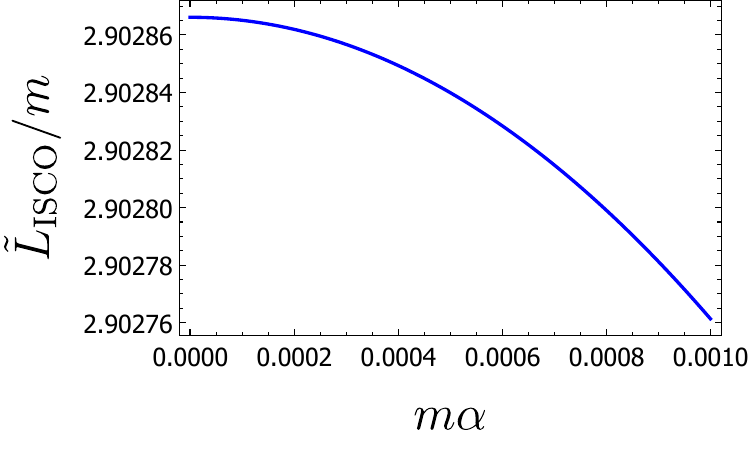}
	\caption{\textit{Top}: The radius of ISCO as a function of acceleration parameter for the case of co-rotating particles. \textit{Bottom}: The angular momentum of the particle in ISCO as a function of acceleration parameter. We have taken $a/m = 0.5$, $\theta = \pi/2$, and $m^2\Lambda = 10^{-52}$ in both panels.}
	\label{fig:isco_vs_alpha}
\end{figure}

\begin{figure}[htp]
	\centering
	\includegraphics[width=0.45\textwidth]{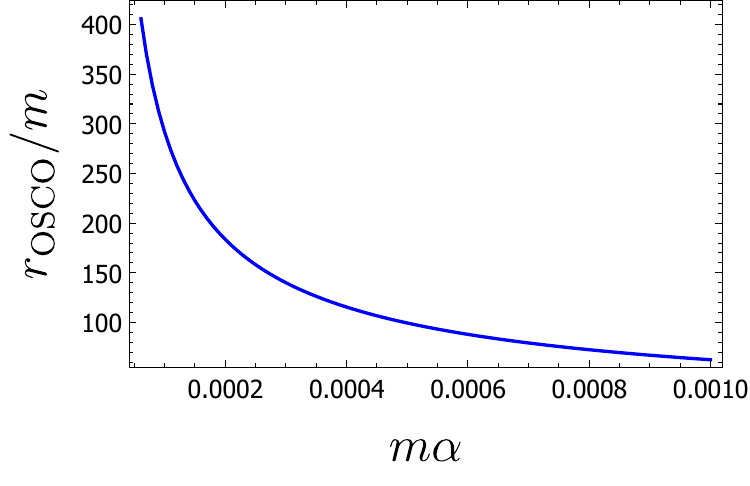}
	\includegraphics[width=0.45\textwidth]{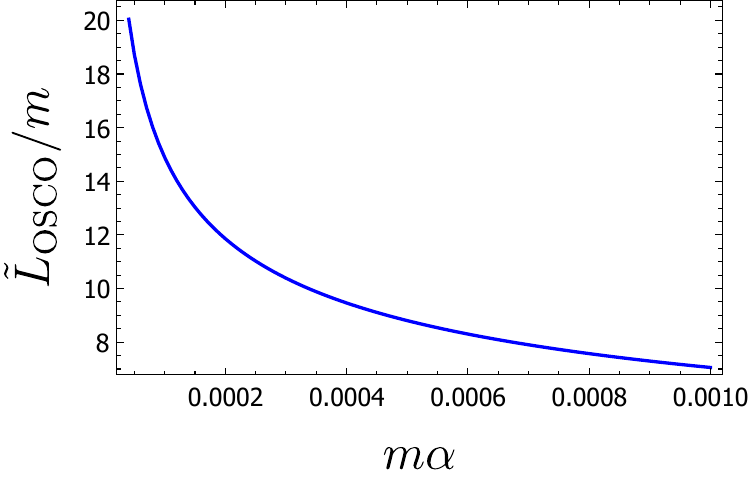}
	\caption{\textit{Top}: The radius of OSCO as a function of acceleration parameter for the case of co-rotating particles. \textit{Bottom}: The angular momentum of the particle in OSCO as a function of acceleration parameter. We have taken $a/m = 0.5$, $\theta = \pi/2$, and $m^2\Lambda = 10^{-52}$ in both panels.}
	\label{fig:osco_vs_alpha}
\end{figure}

Now we would like to study how the radius and angular momentum of inflection points change with the acceleration and rotation parameters. Considering $\tilde{L}>0$ case, we can see in Fig.~\ref{fig:isco_vs_alpha} that the radius of ISCO increases with increasing the acceleration parameter and the angular momentum of particle on this orbit decreases by increasing the acceleration parameter. On the other hand, we see in Fig.~\ref{fig:osco_vs_alpha} that both the radius and angular momentum of OSCO decrease by increasing the acceleration parameter.

\begin{figure}[htp]
	\centering
	\includegraphics[width=0.45\textwidth]{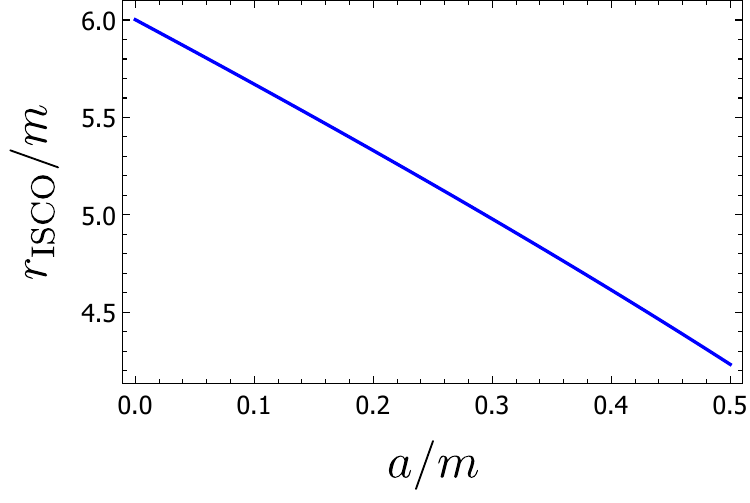}
	\includegraphics[width=0.45\textwidth]{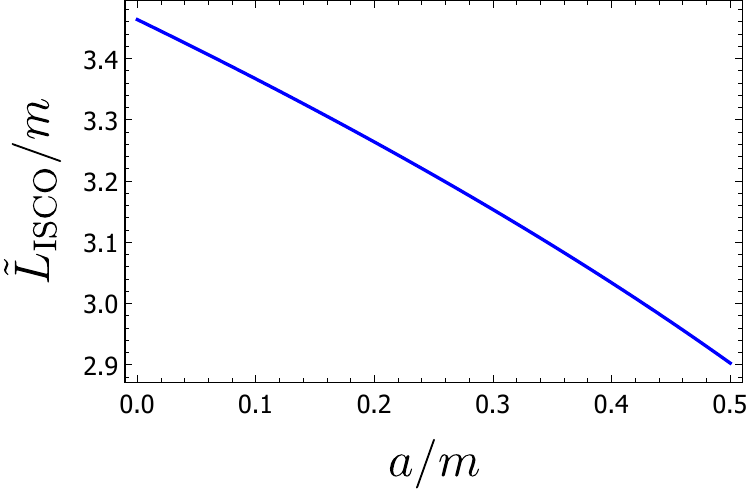}
	\caption{\textit{Top}: The radius of ISCO as a function of rotation parameter for the case of co-rotating particles. \textit{Bottom}: The angular momentum of the particle at ISCO as a function of rotation parameter. We have taken $m\alpha = 10^{-3}$, $\theta = \pi/2$, and $m^2\Lambda = 10^{-52}$ in both panels.}
	\label{fig:isco_vs_a}
\end{figure}

\begin{figure}[htp]
	\centering
	\includegraphics[width=0.45\textwidth]{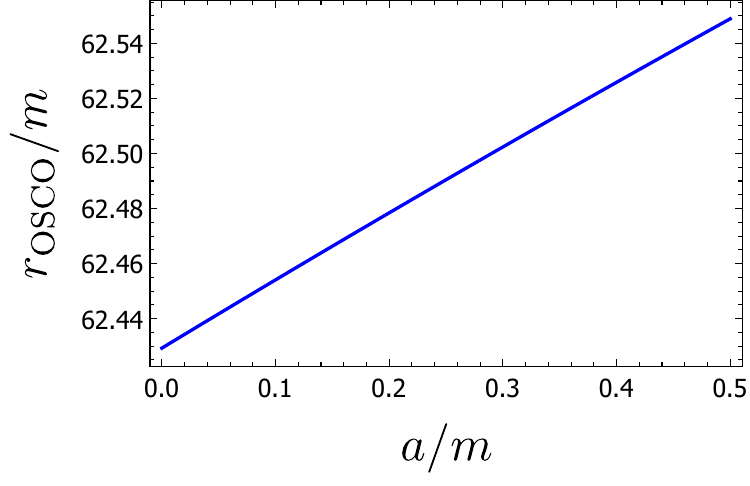}
	\includegraphics[width=0.45\textwidth]{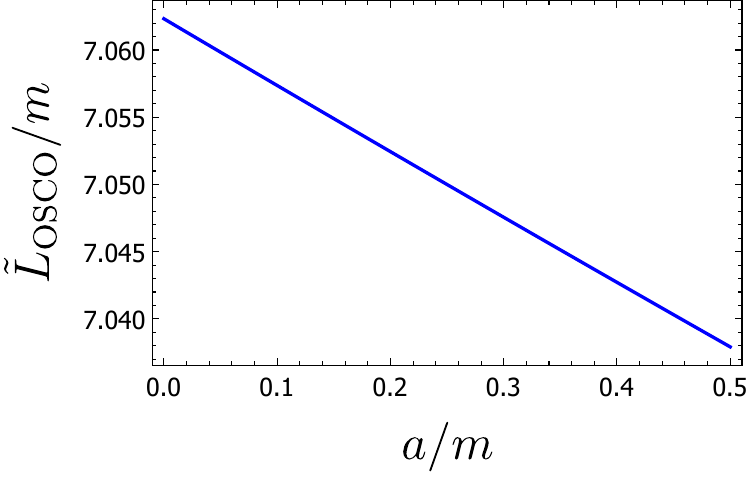}
	\caption{\textit{Top}: The radius of OSCO as a function of rotation parameter for the case of co-rotating particles. \textit{Bottom}: The angular momentum of the particle in OSCO as a function of rotation parameter. We have taken $m\alpha = 10^{-3}$, $\theta = \pi/2$, and $m^2\Lambda = 10^{-52}$ in both panels.}
	\label{fig:osco_vs_a}
\end{figure}

Rotating accelerating black hole has a radius of ISCO which is smaller than that of a non-rotating one. We can see in Fig.~\ref{fig:isco_vs_a} that ISCO radius and angular momentum decrease by increasing the rotation parameter. On the other hand, as can be seen in Fig.~\ref{fig:osco_vs_a}, the radius of OSCO increases by the rotation parameter. However, the angular momentum of particle at OSCO decreases by increasing $a$.

We are not going to present the plots for the counter-rotating cases, however we have found that for these cases the radius of ISCO increases with the acceleration parameter. Also, the absolute value of the angular momentum of particles at ISCO decreases by increasing $\alpha$. The radius of OSCO and the absolute value of angular momentum of particle at this orbit decrease by increasing the acceleration parameter. Also, for the case of $\tilde{L}<0$, the radius of ISCO decreases as the rotation parameter increases and the absolute value of the angular momentum of particles at ISCO increases with the rotation parameter. The radius of OSCO decreases with increasing the rotation parameter, and the absolute value of the angular momentum at OSCO increases with $a$.

Circular motions are of particular interest in the study of accretion disks. For this reason, here we present the equation for angular velocity, energy and angular momentum of particles in circular orbits around the rotating accelerating black holes. These equations will be used in Sec.~\ref{sec:accretion} to find the properties of accretion disks.

\begin{figure}[htp]
	\centering
	\includegraphics[width=0.45\textwidth]{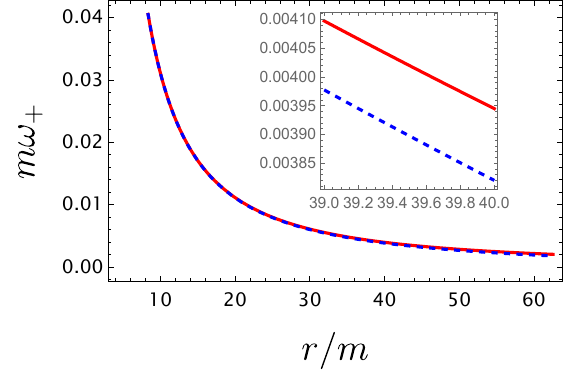}
	\includegraphics[width=0.45\textwidth]{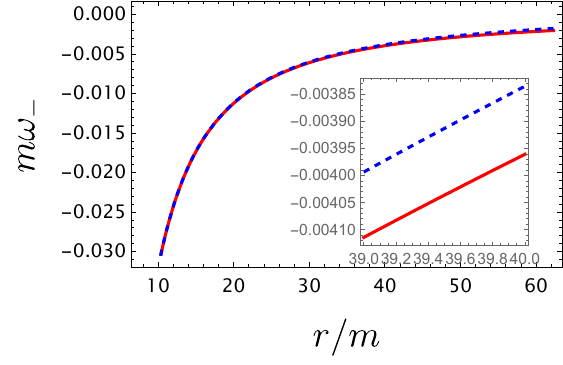}
	\caption{\textit{Top}: Angular velocity of particles on circular orbits co-rotating with the black hole. \textit{Bottom}: Angular velocity of particles on circular orbits counter-rotating around the black hole. In the dashed blue curve we set $m\alpha = 10^{-3}$ and in the red curve we set $\alpha = 0$. We have taken $a/m = 0.5$, $\theta = \pi/2$, and $m^2\Lambda = 10^{-52}$ in both panels. The range of $r$ is from $r_{\rm ISCO}$ to $r_{\rm OSCO}$.}
	\label{fig:omega_pm}
\end{figure}

The geodesic equations can be written as (see \cite{Bambi:2016iip})
\be
g_{\mu\nu,\sigma}\dot{x}^\mu\dot{x}^\nu = 2\frac{d}{d\lambda}(g_{\sigma\mu}\dot{x}^\mu),
\ee
where $\lambda$ is the affine parameter. Here the subscript $,\sigma$ means differentiation with respect to the component $\sigma$. By taking $\sigma = r$, one finds
\be
g_{tt,r}\dot{t}^2+2g_{t\phi,r}\dot{t}\dot{\phi}+g_{\phi\phi,r}\dot{\phi}^2 = 0.
\ee
It is now easy to find the angular velocity $\omega=\dot{\phi}/\dot{t}$ of massive test particle as
\be\label{eqn:omega}
\omega _{\pm}=\frac{-g_{t\phi,r}\pm\sqrt{(g_{t\phi,r})^2-g_{tt,r}g_{\phi\phi,r}}}{g_{\phi\phi,r}},
\ee
where the upper signs are for the case of co-rotating particles and the lower signs are for the counter-rotating particles. In Fig.~\ref{fig:omega_pm} we have plotted the angular velocity of particles co/counter-rotating around the black hole for both accelerating and non-accelerating Kerr black hole. We see that, for co/counter-rotating particle, the absolute value of the angular velocity decreases by increasing radius of the orbit. We also see that for co/counter-rotating particle the absolute value of angular velocity is smaller if the black hole is accelerating.

By using Eq.~\eqref{eqn:lag_geo} and Eq.~\eqref{eqn:omega}, for equatorial circular orbits $\dot{\theta} = \dot{r} = 0$, one finds
\be
\dot{t}_\pm = \frac{\mu}{\sqrt{-g_{tt}-2 g_{t\phi}\omega_{\pm}-g_{\phi\phi}\omega_{\pm}^2}}.
\ee
Therefore, using Eqs.~\eqref{eqn:E} and \eqref{eqn:L_z}, we find the energy and angular momentum of particles on circular orbits as
\ba
\tilde{E}_{\pm}&=&-\frac{g_{tt}+g_{t\phi}\omega_{\pm}}{\sqrt{-g_{tt}-2 g_{t\phi}\omega_{\pm}-g_{\phi\phi}\omega_{\pm}^2}}, \label{eqn:Ec}\\
\tilde{L}_{\pm}&=&\frac{g_{t\phi}+g_{\phi\phi}\omega_{\pm}}{\sqrt{-g_{tt}-2 g_{t\phi}\omega_{\pm}-g_{\phi\phi}\omega_{\pm}^2}}. \label{eqn:Lc}
\ea

\begin{figure}[htp]
	\centering
	\includegraphics[width=0.45\textwidth]{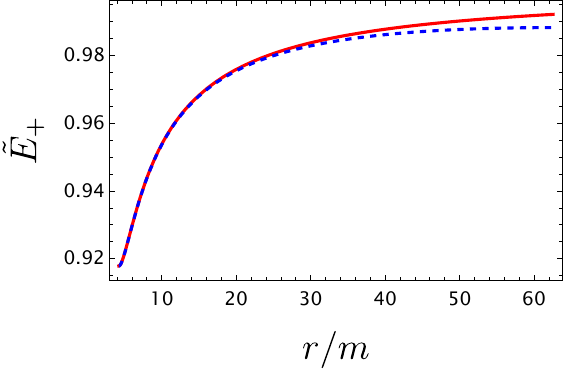}
	\includegraphics[width=0.45\textwidth]{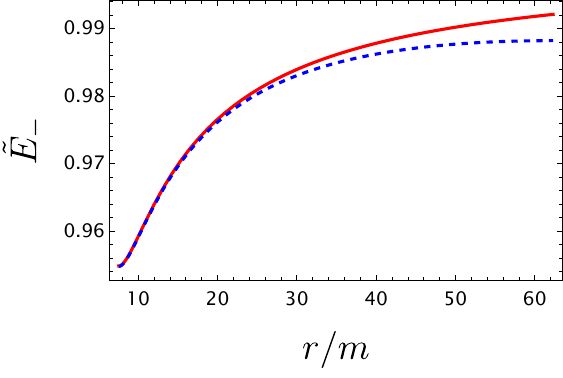}
	\caption{\textit{Top}: Energy of particles on circular orbits co-rotating with the black hole. \textit{Bottom}: Energy of particles on circular orbits counter-rotating around the black hole. In the dashed blue curve we set $m\alpha = 10^{-3}$ and in the red curve we set $\alpha = 0$. We have taken $a/m = 0.5$, $\theta = \pi/2$, and $m^2\Lambda = 10^{-52}$ in both panels. The range of $r$ is from $r_{\rm ISCO}$ to $r_{\rm OSCO}$.}
	\label{fig:E_pm}
\end{figure}

In Fig.~\ref{fig:E_pm} we have plotted the energy of particles co/counter-rotating around the black hole for both accelerating and non-accelerating Kerr black hole. One finds that the energy of the particles increases by increasing the radius of the orbit, whether they are co- or counter-rotating around the black hole. One also find that the energy is smaller if the black hole is accelerating.

\begin{figure}[htp]
	\centering
	\includegraphics[width=0.45\textwidth]{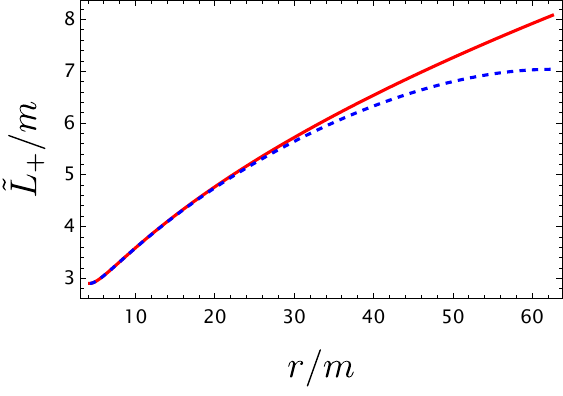}
	\includegraphics[width=0.45\textwidth]{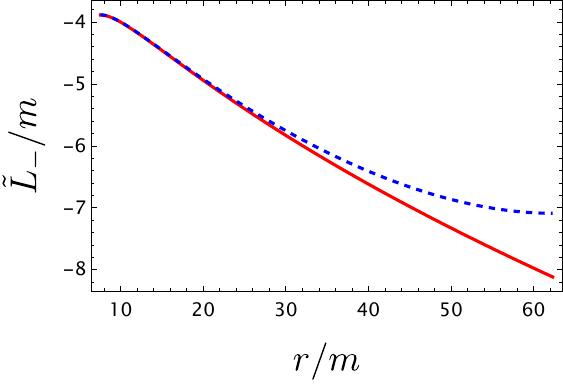}
	\caption{\textit{Top}: Angular momentum of particles on circular orbits co-rotating with the black hole. \textit{Bottom}: Angular momentum of particles on circular orbits counter-rotating around the black hole. In the dashed blue curve we set $m\alpha = 10^{-3}$ and in the red curve we set $\alpha = 0$. We have taken $a/m = 0.5$, $\theta = \pi/2$, and $m^2\Lambda = 10^{-52}$ in both panels. The range of $r$ is from $r_{\rm ISCO}$ to $r_{\rm OSCO}$.}
	\label{fig:L_pm}
\end{figure}

Plots of the angular momentum, Eq. \eqref{eqn:Lc}, are given in Fig.~\ref{fig:L_pm} for both accelerating and non-accelerating black holes. It is obvious that for the co/counter-rotating  particle, the absolute value of the angular momentum increases as the radius gets larger. We can also see that, compared to non-accelerating Kerr black hole, the absolute value of the angular momentum is smaller for accelerating Kerr black hole.

\section{Precession of orbits}\label{sec:precession}

In this section we consider non-circular motion of test particles around rotating accelerating black holes. By using the Eqs.~\eqref{eqn:E} and \eqref{eqn:L_z}, we find from the Lagrangian \eqref{eqn:lag_geo}
\ba\label{eqn:dr}
\left(\frac{dr}{d\phi}\right)^2 &=& (g_{t\phi}^2-g_{tt}g_{\phi\phi})\nn\\ &\times&\frac{g_{\phi\phi}\tilde{E}^2+2g_{t\phi}\tilde{E}\tilde{L}+g_{tt}\tilde{L}^2+g_{tt} g_{\phi\phi}-g_{t\phi}^2}{g_{rr}(g_{t\phi}\tilde{E}+g_{tt}\tilde{L})^2}.\nn\\
\ea
At the turning points which are the perihelion and aphelion of the orbit, the left hand side of the above equation vanishes. We denote the radius of perihelion and aphelion by $r_p$ and $r_a$ respectively. We find the energy and angular momentum per unit rest mass as
\ba
\tilde{E}^{\pm}&=&\sqrt{\frac{\gamma \mp 2\sqrt{\zeta }}{\rho }}, \label{eqn:epm}\\
\tilde{L}^{\pm}&=&\frac{\sqrt{\zeta } \mp \beta}{\xi }\sqrt{\frac{\gamma \mp \sqrt{\zeta }}{\rho}}\label{eqn:lpm},
\ea
where the upper signs are for the case of co-rotating orbits, and the lower signs are for the case of counter-rotating orbits. The auxiliary functions are as follow
\ba
\zeta &=& \left(g_{t\phi}^2(r_a)-g_{tt}(r_a) g_{\phi \phi}(r_a)\right)\left(g_{t \phi}^2(r_p)- g_{tt}(r_p) g_{\phi \phi}(r_p)\right)\nn\\
&\times&(g_{tt}(r_p) g_{t \phi}(r_a)-g_{tt}(r_a) g_{t \phi}^2(r_p)) \left((g_{t \phi}(r_a)-g_{t \phi}(r_p))^2\right.\nn\\
&-&\left.(g_{tt}(r_a)-g_{tt}(r_p)) (g_{\phi \phi}(r_a)-g_{\phi \phi}(r_p))\right),\nn
\ea
\ba
\gamma &=& g_{tt}^2(r_p) \left(g_{tt}(r_a) g_{\phi\phi}(r_a) (g_{\phi\phi}(r_p)-g_{\phi\phi}(r_a))+g_{t\phi}^2(r_a)\right.\nn\\
&\times&\left.(g_{\phi \phi}(r_a)-2 g_{\phi\phi}(r_p))\right)+g_{tt}(r_p) \left(g_{t\phi}^2(r_a) \left(2g_{t\phi}^2(r_p)\right.\right.\nn\\
&-&\left.g_{tt}(r_a) g_{\phi \phi}(r_p)\right)+2 g_{tt}(r_a) g_{t\phi}(r_a) g_{t\phi}(r_p) (g_{\phi\phi}(r_a)\nn\\
&+&g_{\phi\phi}(r_p))- g_{tt}(r_a) g_{t\phi}^2(r_p) g_{\phi\phi}(r_a)+g_{tt}^2(r_a) g_{\phi\phi}(r_p)\nn\\
&\times&(g_{\phi\phi}(r_a)-g_{\phi\phi}(r_p))-2g_{t\phi}^3(r_a)
\left.g_{t\phi}(r_p)\right)+g_{tt}(r_a)\nn\\
&\times&g_{t\phi}^2(r_p) \left(g_{tt}(r_a) (g_{\phi\phi}(r_p)-2 g_{\phi\phi}(r_a))-2g_{t\phi}(r_a)\right.\nn\\
&\times&\left. g_{t\phi}(r_p)+2g_{t\phi}^2(r_a)\right),\nn
\ea
\ba
\rho &=&g_{tt}(r_p) \left(g_{\phi\phi}(r_p) \left(4g_{t\phi}(r_a)^2-2 g_{tt}(r_a)
g_{\phi\phi}(r_a)\right)-4g_{t\phi}(r_a)\right.\nn\\
&\times&\left.g_{t\phi}(r_p) g_{\phi\phi}(r_a)\right)+g_{tt}(r_a) \left(g_{tt}(r_a) g_{\phi\phi}^2(r_p)+4g_{t\phi}^2(r_p)\right.\nn\\
&\times&\left.g_{\phi\phi}(r_a)-4g_{t\phi}(r_a)
g_{t\phi}(r_p) g_{\phi\phi}(r_p)\right)+g_{tt}^2(r_p) g_{\phi\phi}^2(r_a),\nn
\ea
\ba
\beta &=& (g_{tt}(r_p) g_{t\phi}(r_a)-g_{tt}(r_a) g_{t\phi}(r_p)) \left(g_{t\phi}(r_p) \left(- g_{tt}(r_a)\right.\right.\nn\\
&\times&\left.g_{\phi\phi}(r_a)-g_{t\phi}(r_a) g_{t\phi}(r_p)+g_{t\phi}^2(r_a)\right)+ g_{tt}(r_p)\nn\\
&\times&\left.g_{t\phi}(r_a) g_{\phi\phi}(r_p)\right),\nn
\ea
\ba
\xi &=&(g_{tt}(r_p) g_{t\phi}(r_a)-g_{tt}(r_a) g_{t\phi}(r_p))
\left(g_{tt}(r_p) \left(g_{tt}(r_a)\right.\right.\nn\\ &\times&\left.\left.(g_{\phi\phi}(r_p)-g_{\phi\phi}(r_a))+g_{t\phi}(r_a)^2\right)-g_{tt}(r_a) g_{t\phi}(r_p)^2\right).\nn
\ea

To find the precession we use the following formula along with Eq.~\eqref{eqn:dr}
\ba
	\Delta\phi&=&2\left[\phi(r_a)-\phi(r_p)\right]-2(1+2\alpha m)^{-1}\pi \nn\\
	=&2&\int_{r_p}^{r_a}\left(\frac{g_{rr}(g_{t\phi}\tilde{E}+g_{tt}\tilde{L})^2}{(g_{t\phi}^2- g_{tt}g_{\phi\phi})\delta}\right)^{1/2}-2(1+2\alpha m)^{-1}\pi, \nn\\
\ea
where
\be
\delta=g_{\phi\phi}\tilde{E}^2+2g_{t\phi}\tilde{E}\tilde{L}+g_{tt}\tilde{L}^2+g_{tt} g_{\phi\phi}-g_{t\phi}^2. \nn
\ee
For the case of co-rotating particles we use $\tilde{E}=\tilde{E}^+$ and  $\tilde{L}=\tilde{L}^+$ and for counter-rotating case we use $\tilde{E}=\tilde{E}^-$ and  $\tilde{L}=\tilde{L}^-$. Also, to find $r_p$ and $r_a$ we use the data of S2 star orbit around Sgr A*. The semi-major axis and eccentricity of the orbit are measured to be $b=1.543\times10^{14} {\rm m}$ and $e=0.88$ \cite{eisenhauer2003geometric}, respectively. These quantities are related to perihelion and aphelion radius through $r_p=(1-e)b$ and $r_a=(1+e)b$.

\begin{figure}[htp]
	\centering
	\includegraphics[width=0.45\textwidth]{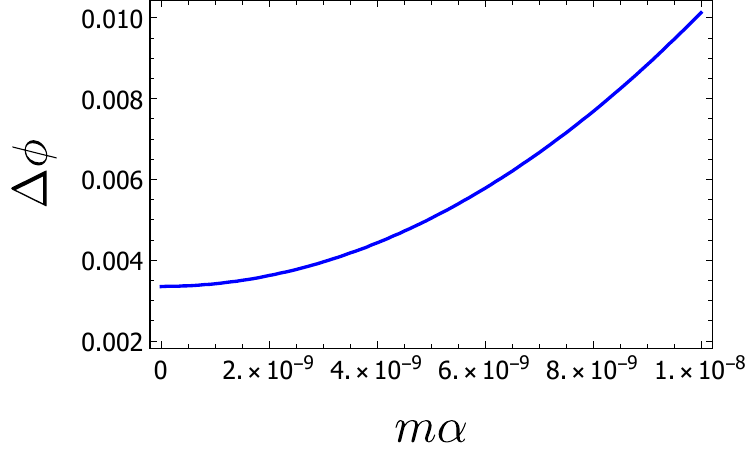}
	\includegraphics[width=0.45\textwidth]{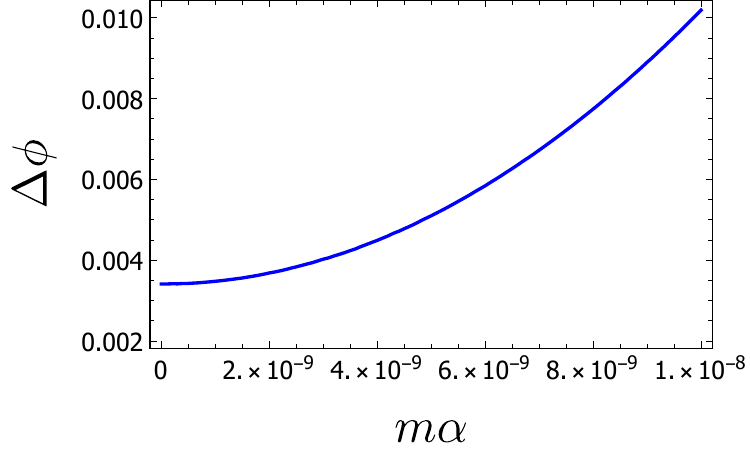}
	\caption{\textit{Top}: The precession of orbit (in radians) of co-rotating particles as a function of the acceleration parameter. \textit{Bottom}: The precession of orbit of counter-rotating particles as a function of the acceleration parameter. We have taken $a=0.5$, $\theta = \pi/2$, and $m^2\Lambda = 10^{-52}$ in both panels. We have used the semi-major axis and eccentricity of the orbit of S2 around Sgr A*.}
	\label{fig:pre}
\end{figure}

In the top panel of Fig.~\ref{fig:pre} we have plotted the value of precession of perihelion for particles co-rotating with the accelerating rotating black hole. In the bottom panel of Fig.~\ref{fig:pre}, the precession is plotted for counter-rotating particles. We see that for both cases the precession is positive. Not only that, for the case of counter-rotating particles the precession is about $10^{-4}$ radians larger. This is true even for the case of non-accelerating black holes. However, to the best of our knowledge, it has not been reported in the literature of rotating black holes earlier.

\section{Accretion disks}\label{sec:accretion}

In this section we study the thermal spectrum of geometrically thin optically thick accretion disks around rotating accelerating black holes. We take this disk to be in equatorial plane of the black hole. In fact, if the disk around a rotating black hole is tilted with respect to the equatorial plane, the inertial frame dragging would drive it to the equatorial plane \cite{Bardeen:1975zz}. We also assume that the accretion rate is low and the pressure and self gravity of disk are negligible. In such a case we can describe the thin disk as consist of particles with circular geodesic orbits in black hole spacetime \cite{Novikov:1973}.

The inner edge of the accretion disk is determined by ISCO. (Particles at $r < r_{\rm ISCO}$ would fall into the black hole in a very short time --- more about such plunging orbits can be found in \cite{Mummery:2023tgh} and the references therein.) In the presence of acceleration parameter and/or a positive cosmological constant, we have OSCO along with ISCO. We take the outer edge of the disk to be at OSCO. The presence of an outer edge has been already seen for (non-accelerating) black holes in de Sitter background \cite{Stuchlik:1999qk,Stuchlik:2002tj}. We note that there is no stable circular orbits in regions $r < r_{\rm ISCO}$ or $r > r_{\rm OSCO}$. This is an important feature of relativistic theory of gravity.

Accretion disks emit radiation due to frictional heating and gravitational potential energy. The radiation from accretion disks is often approximated by a black body spectrum, which is a function of the disk temperature and radius. The disk temperature decreases with increasing radius, so the inner regions of the disk emit more high-energy radiation, such as X-rays, while the outer regions emit more low-energy radiation, such as infrared \cite{Li:2004aq}. The shape of the black body spectrum also depends on the spin and mass of the black hole \cite{Reynolds:2013qqa}. By observing the radiation from accretion disks, we can infer the properties of the black holes and test the predictions of general relativity.

Here we are interested in electromagnetic radiation of the accretion disk. The amount of energy per unit area per unit time that is emitted by the disk is known as the radiative flux which is usually calculated by solving the equations of radiative transfer and energy conservation in the disk \cite{Narzilloev:2022avv}. For geometrically thin, optically thick accretion disk in a steady state the radiative flux can be expressed as
\ba
\mathcal{F}_{\pm}(r) &=& -\frac{\dot{m}\omega_{\pm,r}}{4\pi\sqrt{-g}(\tilde{E}_{\pm}-\omega_{\pm}\tilde{L}_{\pm})^2} \nn\\ &\times&\int_{r_{\rm ISCO}}^{r}(\tilde{E}_{\pm}-\omega_{\pm}\tilde{L}_{\pm})\tilde{L}_{\pm,\bar{r}}d\bar{r},
\ea
where $\dot{m}$ is the mass accretion rate which we take to be constant and $g$ is the determinant of the metric of the 2+1 dimensional subspace $(t, r, \phi)$ and is given by $\sqrt{-g} = \sqrt{-g_{rr}\left(g_{tt}g_{\phi\phi}-g_{t\phi}^2\right)}$.

\begin{figure}[htp]
	\centering
	\includegraphics[width=0.45\textwidth]{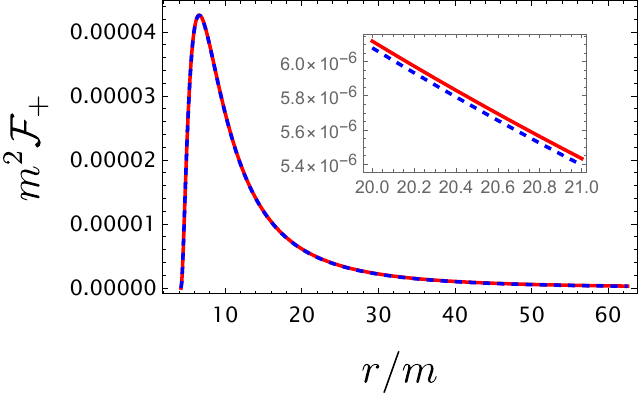}
	\includegraphics[width=0.45\textwidth]{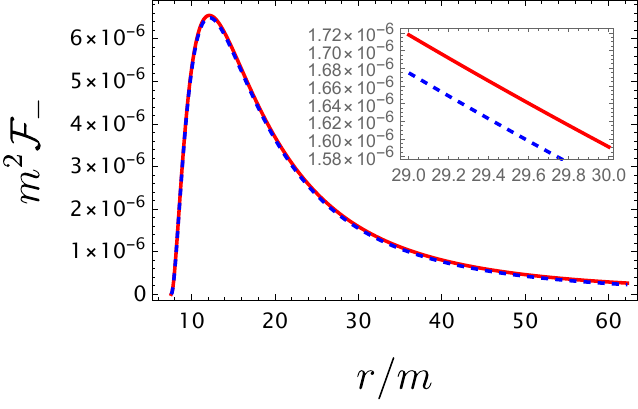}
	\caption{\textit{Top}: Radiative flux of a disk co-rotating with the black hole. \textit{Bottom}: Radiative flux of a disk counter-rotating around the black hole. In the dashed blue curve we set $m\alpha = 10^{-3}$ and in the red curve we set $\alpha = 0$. We have taken $a/m = 0.5$, $\theta = \pi/2$, and $m^2\Lambda = 10^{-52}$ in both panels. The range of $r$ is from $r_{\rm ISCO}$ to $r_{\rm OSCO}$.}
	\label{fig:sqf}
\end{figure}

In Fig.~\ref{fig:sqf} we have plotted the radiative flux of a co-rotating (top panel) and counter-rotating (bottom panel) disk around a rotating black hole which can be accelerating or not. In this figure and in the rest of this section we take $\dot{m}=1$, or equivalently we measure the radiative flux in units of accretion rate. We see that the radiative flux is larger if the disk is co-rotating with the black hole. Also, for both co-rotating and counter-rotating cases the flux is lower for accelerating black hole.

\begin{figure}[htp]
	\centering
	\includegraphics[width=0.45\textwidth]{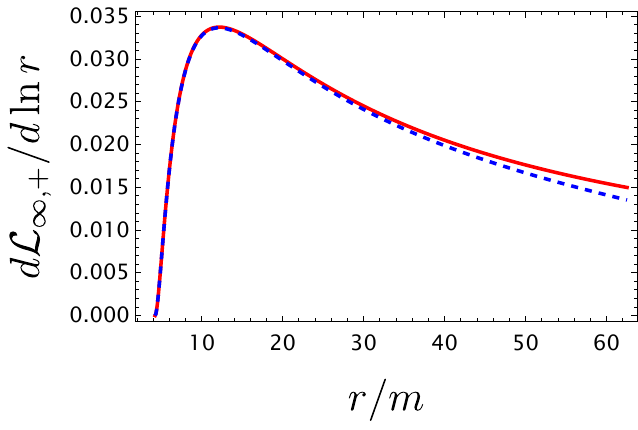}
	\includegraphics[width=0.45\textwidth]{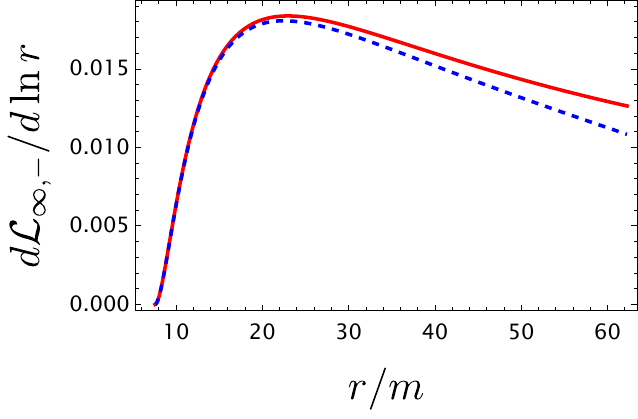}
	\caption{\textit{Top}: Differential luminosity of a disk co-rotating with the black hole. \textit{Bottom}: Differential luminosity of a disk counter-rotating around the black hole. In the dashed blue curve we set $m\alpha = 10^{-3}$ and in the red curve we set $\alpha = 0$. We have taken $a/m = 0.5$, $\theta = \pi/2$, and $m^2\Lambda = 10^{-52}$ in both panels. The range of $r$ is from $r_{\rm ISCO}$ to $r_{\rm OSCO}$.}
	\label{fig:sqdl}
\end{figure}

The radiative flux is a local quantity measured in accretion disk rest frame. The energy per unit time received by an observer at infinity is denoted by the luminosity $\mathcal{L}_{\infty}$ and satisfy the following relation which is known as the differential luminosity \cite{Novikov:1973,Page:1974he}
\be
\frac{d\mathcal{L}_{\infty,\pm}}{d \ln r} = 4\pi r \sqrt{-g}\tilde{E}_{\pm} \mathcal{F}_\pm(r).
\ee
Plots of the above quantity for both co-rotating (top panel) and counter-rotating (bottom panel) disks are presented in Fig.~\ref{fig:sqdl}. We find that this quantity is larger for co-rotating disk and it is smaller if the black hole is accelerating.

\begin{figure}[htp]
	\centering
	\includegraphics[width=0.45\textwidth]{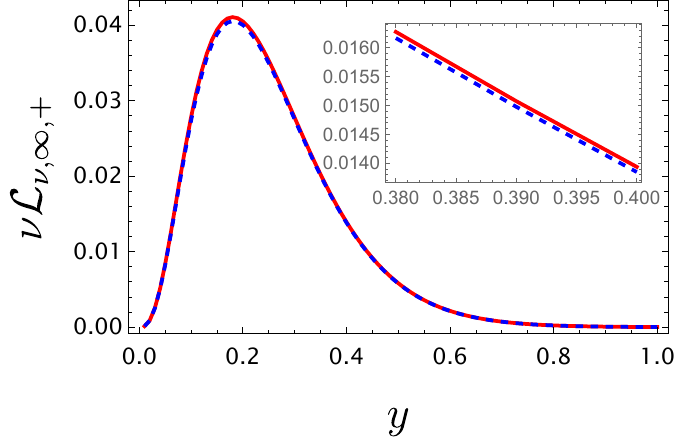}
	\includegraphics[width=0.45\textwidth]{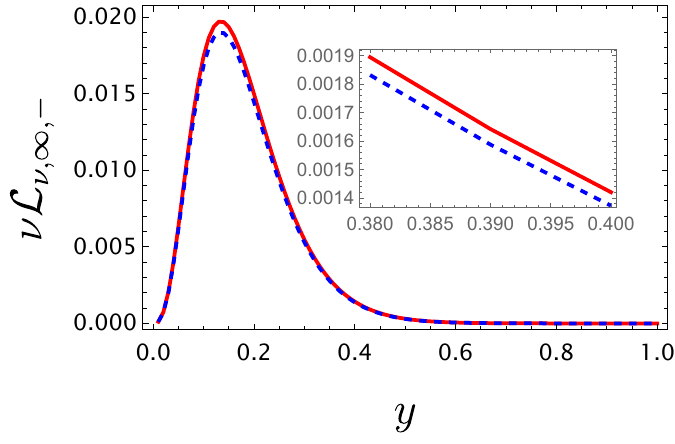}
	\caption{\textit{Top}: Spectral luminosity distribution of a disk co-rotating with the black hole. \textit{Bottom}: Spectral luminosity distribution of a disk counter-rotating around the black hole. In the dashed blue curve we set $m\alpha = 10^{-3}$ and in the red curve we set $\alpha = 0$. We have taken $a/m = 0.5$, $\theta = \pi/2$, and $m^2\Lambda = 10^{-52}$ in both panels. The range of $r$ is from $r_{\rm ISCO}$ to $r_{\rm OSCO}$.}
	\label{fig:sqnul}
\end{figure}

The quantity of most observational importance is the spectral luminosity distribution $\mathcal{L}_{\nu, \infty}$, which is the amount of energy per unit frequency that is emitted by the disk as electromagnetic radiation over the entire spectrum. By assuming the emission of the disk to be similar to black body radiation, we find the spectral luminosity as \cite{Boshkayev:2020kle}
\be
\nu\mathcal{L}_{\nu, \infty,\pm} = \frac{60}{\pi^3} \int_{r_{\rm ISCO}}^{r_{\rm OSCO}}\frac{\sqrt{-g}\tilde{E}_\pm (u_\pm^t y)^4}{m^2\left(\exp\left[u_\pm^t y/(m^2\mathcal{F}_\pm)^{1/4}\right]-1\right)}dr,
\ee
where
\ba
u_\pm^t(r) &=& \frac{1}{\sqrt{-g_{tt}-2\omega_{\pm}g_{t\phi}-\omega_{\pm}^2g_{\phi\phi}}},\\
y &=& \frac{h\nu}{kT_*},\\
T_*^4 &=& \frac{\dot{m}}{4\pi\sigma m^2},
\ea
with $h$, $k$, and $\sigma$ denoting Planck, Boltzmann, and Stefan–Boltzmann constant constants, respectively. The plots of spectral luminosity distribution are presented in Fig.~\ref{fig:sqnul}, for the case of co-rotating disk (top panel) and counter-rotating disk (bottom panel). It is obvious that the spectral luminosity is larger if the disk co-rotates with the black hole. It is also larger for non-accelerating black hole.

\section{Concluding remarks}\label{sec:con}

In this paper, we investigated the motion of test particles and the properties of accretion disks in rotating, accelerating black hole spacetimes with a non-zero cosmological constant.

Specifically, we examined how both rotation and acceleration affect the dynamics of massive particles orbiting around the quoted black hole configurations.

To this end, we first selected the Plebański-Demiański spacetime and thus derived the geodesic equations, working out the characteristics of circular orbits, including ISCOs and OSCOs.

Our findings definitely indicated the great impact of the acceleration parameter on the radii and angular momenta of these orbits.

Precisely, the inner and outer edges of accretion disk depend on the rotation and acceleration parameters and the sense of rotation. Not only the shape of the accretion disks, but the spectral quantities of the disk depend on these parameters. In particular we have shown that the acceleration decreases the radiative flux and luminosity of the disk.

We further investigated the thermal spectra of geometrically thin, optically thick accretion disks. As background model, we employed the widely-consolidated  Novikov-Thorne scenario and, there, by analyzing both co-rotating and counter-rotating disks, we demonstrated that the radiative flux and luminosity are significantly higher for co-rotating disks.

Further, we compared the spectral features and luminosity of accretion disks around accelerating black holes with those around non-accelerating Kerr black holes, finding that acceleration generally reduces the luminosity and flux of emitted light.

Our results therefore provide very useful hints to further clarify the interplay of black hole acceleration and rotation with the dynamics of the surrounding accretion disk and also the role that acceleration may play in possible future experiments that could reveal the observational signature of such objects.

Accordingly, as future steps, we can consider,

\begin{itemize}
	\item[-] the investigation of  motion of spinning test particles around rotating, accelerating black holes;
	\item[-] including self-force effects on test particles, leading to more accurate description of their motion, especially in strong gravitational fields near the black hole, namely including possible particle backreaction  \cite{Poisson:2011nh};
	\item[-] computing refined numerical simulations of particle motion and accretion disk dynamics in the presence of both rotation and acceleration,  exploring regimes that are difficult to tackle analytically.
\end{itemize}

Last but not least, we leave open the chance of finding plausible observational signatures with next year observations, i.e. identifying potential observational signatures of rotating, accelerating black holes, namely clarifying the possible role of acceleration in the astrophysics of compact objects.


\begin{thebibliography}{67}%
\makeatletter
\providecommand \@ifxundefined [1]{%
 \@ifx{#1\undefined}
}%
\providecommand \@ifnum [1]{%
 \ifnum #1\expandafter \@firstoftwo
 \else \expandafter \@secondoftwo
 \fi
}%
\providecommand \@ifx [1]{%
 \ifx #1\expandafter \@firstoftwo
 \else \expandafter \@secondoftwo
 \fi
}%
\providecommand \natexlab [1]{#1}%
\providecommand \enquote  [1]{``#1''}%
\providecommand \bibnamefont  [1]{#1}%
\providecommand \bibfnamefont [1]{#1}%
\providecommand \citenamefont [1]{#1}%
\providecommand \href@noop [0]{\@secondoftwo}%
\providecommand \href [0]{\begingroup \@sanitize@url \@href}%
\providecommand \@href[1]{\@@startlink{#1}\@@href}%
\providecommand \@@href[1]{\endgroup#1\@@endlink}%
\providecommand \@sanitize@url [0]{\catcode `\\12\catcode `\$12\catcode `\&12\catcode `\#12\catcode `\^12\catcode `\_12\catcode `\%12\relax}%
\providecommand \@@startlink[1]{}%
\providecommand \@@endlink[0]{}%
\providecommand \url  [0]{\begingroup\@sanitize@url \@url }%
\providecommand \@url [1]{\endgroup\@href {#1}{\urlprefix }}%
\providecommand \urlprefix  [0]{URL }%
\providecommand \Eprint [0]{\href }%
\providecommand \doibase [0]{http://dx.doi.org/}%
\providecommand \selectlanguage [0]{\@gobble}%
\providecommand \bibinfo  [0]{\@secondoftwo}%
\providecommand \bibfield  [0]{\@secondoftwo}%
\providecommand \translation [1]{[#1]}%
\providecommand \BibitemOpen [0]{}%
\providecommand \bibitemStop [0]{}%
\providecommand \bibitemNoStop [0]{.\EOS\space}%
\providecommand \EOS [0]{\spacefactor3000\relax}%
\providecommand \BibitemShut  [1]{\csname bibitem#1\endcsname}%
\let\auto@bib@innerbib\@empty
\bibitem [{\citenamefont {Damour}\ \emph {et~al.}(2003)\citenamefont {Damour}, \citenamefont {Iyer}, \citenamefont {Jaranowski},\ and\ \citenamefont {Sathyaprakash}}]{Damour:2002vi}%
  \BibitemOpen
  \bibfield  {author} {\bibinfo {author} {\bibfnamefont {T.}~\bibnamefont {Damour}}, \bibinfo {author} {\bibfnamefont {B.~R.}\ \bibnamefont {Iyer}}, \bibinfo {author} {\bibfnamefont {P.}~\bibnamefont {Jaranowski}}, \ and\ \bibinfo {author} {\bibfnamefont {B.~S.}\ \bibnamefont {Sathyaprakash}},\ }\href {\doibase 10.1103/PhysRevD.67.064028} {\bibfield  {journal} {\bibinfo  {journal} {Phys. Rev. D}\ }\textbf {\bibinfo {volume} {67}},\ \bibinfo {pages} {064028} (\bibinfo {year} {2003})},\ \Eprint {http://arxiv.org/abs/gr-qc/0211041} {arXiv:gr-qc/0211041} \BibitemShut {NoStop}%
\bibitem [{\citenamefont {Ustyugova}\ \emph {et~al.}(2000)\citenamefont {Ustyugova}, \citenamefont {Lovelace}, \citenamefont {Romanova}, \citenamefont {Li},\ and\ \citenamefont {Colgate}}]{ustyugova2000poynting}%
  \BibitemOpen
  \bibfield  {author} {\bibinfo {author} {\bibfnamefont {G.~V.}\ \bibnamefont {Ustyugova}}, \bibinfo {author} {\bibfnamefont {R.~V.~E.}\ \bibnamefont {Lovelace}}, \bibinfo {author} {\bibfnamefont {M.~M.}\ \bibnamefont {Romanova}}, \bibinfo {author} {\bibfnamefont {H.}~\bibnamefont {Li}}, \ and\ \bibinfo {author} {\bibfnamefont {S.~A.}\ \bibnamefont {Colgate}},\ }\href {\doibase 10.1086/312890} {\bibfield  {journal} {\bibinfo  {journal} {Astrophys. J.}\ }\textbf {\bibinfo {volume} {541}},\ \bibinfo {pages} {:L21} (\bibinfo {year} {2000})}\BibitemShut {NoStop}%
\bibitem [{\citenamefont {Shibata}\ and\ \citenamefont {Sasaki}(1999)}]{Shibata:1999zs}%
  \BibitemOpen
  \bibfield  {author} {\bibinfo {author} {\bibfnamefont {M.}~\bibnamefont {Shibata}}\ and\ \bibinfo {author} {\bibfnamefont {M.}~\bibnamefont {Sasaki}},\ }\href {\doibase 10.1103/PhysRevD.60.084002} {\bibfield  {journal} {\bibinfo  {journal} {Phys. Rev. D}\ }\textbf {\bibinfo {volume} {60}},\ \bibinfo {pages} {084002} (\bibinfo {year} {1999})},\ \Eprint {http://arxiv.org/abs/gr-qc/9905064} {arXiv:gr-qc/9905064} \BibitemShut {NoStop}%
\bibitem [{\citenamefont {Jaklitsch}\ \emph {et~al.}(1989)\citenamefont {Jaklitsch}, \citenamefont {Hellaby},\ and\ \citenamefont {Matravers}}]{jaklitsch1989}%
  \BibitemOpen
  \bibfield  {author} {\bibinfo {author} {\bibfnamefont {M.~J.}\ \bibnamefont {Jaklitsch}}, \bibinfo {author} {\bibfnamefont {C.}~\bibnamefont {Hellaby}}, \ and\ \bibinfo {author} {\bibfnamefont {D.~R.}\ \bibnamefont {Matravers}},\ }\href {\doibase 10.1007/BF00769865} {\bibfield  {journal} {\bibinfo  {journal} {Gen. Rel. Grav.}\ }\textbf {\bibinfo {volume} {21}},\ \bibinfo {pages} {941} (\bibinfo {year} {1989})}\BibitemShut {NoStop}%
\bibitem [{\citenamefont {Cruz}\ \emph {et~al.}(2005)\citenamefont {Cruz}, \citenamefont {Olivares},\ and\ \citenamefont {Villanueva}}]{Cruz:2004ts}%
  \BibitemOpen
  \bibfield  {author} {\bibinfo {author} {\bibfnamefont {N.}~\bibnamefont {Cruz}}, \bibinfo {author} {\bibfnamefont {M.}~\bibnamefont {Olivares}}, \ and\ \bibinfo {author} {\bibfnamefont {J.~R.}\ \bibnamefont {Villanueva}},\ }\href {\doibase 10.1088/0264-9381/22/6/016} {\bibfield  {journal} {\bibinfo  {journal} {Class. Quant. Grav.}\ }\textbf {\bibinfo {volume} {22}},\ \bibinfo {pages} {1167} (\bibinfo {year} {2005})},\ \Eprint {http://arxiv.org/abs/gr-qc/0408016} {arXiv:gr-qc/0408016} \BibitemShut {NoStop}%
\bibitem [{\citenamefont {Stuchlik}\ and\ \citenamefont {Hledik}(2002)}]{Stuchlik:2002tj}%
  \BibitemOpen
  \bibfield  {author} {\bibinfo {author} {\bibfnamefont {Z.}~\bibnamefont {Stuchlik}}\ and\ \bibinfo {author} {\bibfnamefont {S.}~\bibnamefont {Hledik}},\ }\href@noop {} {\bibfield  {journal} {\bibinfo  {journal} {Acta Phys. Slov.}\ }\textbf {\bibinfo {volume} {52}},\ \bibinfo {pages} {363} (\bibinfo {year} {2002})},\ \Eprint {http://arxiv.org/abs/0803.2685} {arXiv:0803.2685 [gr-qc]} \BibitemShut {NoStop}%
\bibitem [{\citenamefont {Pugliese}\ \emph {et~al.}(2011)\citenamefont {Pugliese}, \citenamefont {Quevedo},\ and\ \citenamefont {Ruffini}}]{Pugliese:2011py}%
  \BibitemOpen
  \bibfield  {author} {\bibinfo {author} {\bibfnamefont {D.}~\bibnamefont {Pugliese}}, \bibinfo {author} {\bibfnamefont {H.}~\bibnamefont {Quevedo}}, \ and\ \bibinfo {author} {\bibfnamefont {R.}~\bibnamefont {Ruffini}},\ }\href {\doibase 10.1103/PhysRevD.83.104052} {\bibfield  {journal} {\bibinfo  {journal} {Phys. Rev. D}\ }\textbf {\bibinfo {volume} {83}},\ \bibinfo {pages} {104052} (\bibinfo {year} {2011})},\ \Eprint {http://arxiv.org/abs/1103.1807} {arXiv:1103.1807 [gr-qc]} \BibitemShut {NoStop}%
\bibitem [{\citenamefont {Fayos}\ and\ \citenamefont {Teijon}(2008)}]{Fayos:2007ks}%
  \BibitemOpen
  \bibfield  {author} {\bibinfo {author} {\bibfnamefont {F.}~\bibnamefont {Fayos}}\ and\ \bibinfo {author} {\bibfnamefont {C.}~\bibnamefont {Teijon}},\ }\href {\doibase 10.1007/s10714-008-0629-1} {\bibfield  {journal} {\bibinfo  {journal} {Gen. Rel. Grav.}\ }\textbf {\bibinfo {volume} {40}},\ \bibinfo {pages} {2433} (\bibinfo {year} {2008})},\ \Eprint {http://arxiv.org/abs/0706.1455} {arXiv:0706.1455 [gr-qc]} \BibitemShut {NoStop}%
\bibitem [{\citenamefont {Stuchlik}\ and\ \citenamefont {Slany}(2004)}]{Stuchlik:2003dt}%
  \BibitemOpen
  \bibfield  {author} {\bibinfo {author} {\bibfnamefont {Z.}~\bibnamefont {Stuchlik}}\ and\ \bibinfo {author} {\bibfnamefont {P.}~\bibnamefont {Slany}},\ }\href {\doibase 10.1103/PhysRevD.69.064001} {\bibfield  {journal} {\bibinfo  {journal} {Phys. Rev. D}\ }\textbf {\bibinfo {volume} {69}},\ \bibinfo {pages} {064001} (\bibinfo {year} {2004})},\ \Eprint {http://arxiv.org/abs/gr-qc/0307049} {arXiv:gr-qc/0307049} \BibitemShut {NoStop}%
\bibitem [{\citenamefont {Soroushfar}\ \emph {et~al.}(2016)\citenamefont {Soroushfar}, \citenamefont {Saffari}, \citenamefont {Kazempour}, \citenamefont {Grunau},\ and\ \citenamefont {Kunz}}]{Soroushfar:2016esy}%
  \BibitemOpen
  \bibfield  {author} {\bibinfo {author} {\bibfnamefont {S.}~\bibnamefont {Soroushfar}}, \bibinfo {author} {\bibfnamefont {R.}~\bibnamefont {Saffari}}, \bibinfo {author} {\bibfnamefont {S.}~\bibnamefont {Kazempour}}, \bibinfo {author} {\bibfnamefont {S.}~\bibnamefont {Grunau}}, \ and\ \bibinfo {author} {\bibfnamefont {J.}~\bibnamefont {Kunz}},\ }\href {\doibase 10.1103/PhysRevD.94.024052} {\bibfield  {journal} {\bibinfo  {journal} {Phys. Rev. D}\ }\textbf {\bibinfo {volume} {94}},\ \bibinfo {pages} {024052} (\bibinfo {year} {2016})},\ \Eprint {http://arxiv.org/abs/1605.08976} {arXiv:1605.08976 [gr-qc]} \BibitemShut {NoStop}%
\bibitem [{\citenamefont {Das}\ \emph {et~al.}(2021)\citenamefont {Das}, \citenamefont {Saha},\ and\ \citenamefont {Gangopadhyay}}]{Das:2020yxw}%
  \BibitemOpen
  \bibfield  {author} {\bibinfo {author} {\bibfnamefont {A.}~\bibnamefont {Das}}, \bibinfo {author} {\bibfnamefont {A.}~\bibnamefont {Saha}}, \ and\ \bibinfo {author} {\bibfnamefont {S.}~\bibnamefont {Gangopadhyay}},\ }\href {\doibase 10.1088/1361-6382/abd95b} {\bibfield  {journal} {\bibinfo  {journal} {Class. Quant. Grav.}\ }\textbf {\bibinfo {volume} {38}},\ \bibinfo {pages} {065015} (\bibinfo {year} {2021})},\ \Eprint {http://arxiv.org/abs/2009.03644} {arXiv:2009.03644 [gr-qc]} \BibitemShut {NoStop}%
\bibitem [{\citenamefont {Wilkins}(1972)}]{Wilkins:1972rs}%
  \BibitemOpen
  \bibfield  {author} {\bibinfo {author} {\bibfnamefont {D.~C.}\ \bibnamefont {Wilkins}},\ }\href {\doibase 10.1103/PhysRevD.5.814} {\bibfield  {journal} {\bibinfo  {journal} {Phys. Rev. D}\ }\textbf {\bibinfo {volume} {5}},\ \bibinfo {pages} {814} (\bibinfo {year} {1972})}\BibitemShut {NoStop}%
\bibitem [{\citenamefont {Teo}(2021)}]{Teo:2020sey}%
  \BibitemOpen
  \bibfield  {author} {\bibinfo {author} {\bibfnamefont {E.}~\bibnamefont {Teo}},\ }\href {\doibase 10.1007/s10714-020-02782-z} {\bibfield  {journal} {\bibinfo  {journal} {Gen. Rel. Grav.}\ }\textbf {\bibinfo {volume} {53}},\ \bibinfo {pages} {10} (\bibinfo {year} {2021})},\ \Eprint {http://arxiv.org/abs/2007.04022} {arXiv:2007.04022 [gr-qc]} \BibitemShut {NoStop}%
\bibitem [{\citenamefont {Luongo}\ and\ \citenamefont {Quevedo}(2014)}]{Luongo:2014qoa}%
  \BibitemOpen
  \bibfield  {author} {\bibinfo {author} {\bibfnamefont {O.}~\bibnamefont {Luongo}}\ and\ \bibinfo {author} {\bibfnamefont {H.}~\bibnamefont {Quevedo}},\ }\href {\doibase 10.1103/PhysRevD.90.084032} {\bibfield  {journal} {\bibinfo  {journal} {Phys. Rev. D}\ }\textbf {\bibinfo {volume} {90}},\ \bibinfo {pages} {084032} (\bibinfo {year} {2014})},\ \Eprint {http://arxiv.org/abs/1407.1530} {arXiv:1407.1530 [gr-qc]} \BibitemShut {NoStop}%
\bibitem [{\citenamefont {Luongo}\ and\ \citenamefont {Quevedo}(2018)}]{Luongo:2015zaa}%
  \BibitemOpen
  \bibfield  {author} {\bibinfo {author} {\bibfnamefont {O.}~\bibnamefont {Luongo}}\ and\ \bibinfo {author} {\bibfnamefont {H.}~\bibnamefont {Quevedo}},\ }\href {\doibase 10.1007/s10701-017-0125-0} {\bibfield  {journal} {\bibinfo  {journal} {Found. Phys.}\ }\textbf {\bibinfo {volume} {48}},\ \bibinfo {pages} {17} (\bibinfo {year} {2018})},\ \Eprint {http://arxiv.org/abs/1507.06446} {arXiv:1507.06446 [gr-qc]} \BibitemShut {NoStop}%
\bibitem [{\citenamefont {Luongo}\ and\ \citenamefont {Quevedo}(2024)}]{Luongo:2023aib}%
  \BibitemOpen
  \bibfield  {author} {\bibinfo {author} {\bibfnamefont {O.}~\bibnamefont {Luongo}}\ and\ \bibinfo {author} {\bibfnamefont {H.}~\bibnamefont {Quevedo}},\ }\href {\doibase 10.1088/1361-6382/ad4ae4} {\bibfield  {journal} {\bibinfo  {journal} {Class. Quant. Grav.}\ }\textbf {\bibinfo {volume} {41}},\ \bibinfo {pages} {125011} (\bibinfo {year} {2024})},\ \Eprint {http://arxiv.org/abs/2305.11185} {arXiv:2305.11185 [gr-qc]} \BibitemShut {NoStop}%
\bibitem [{\citenamefont {Comp\`ere}\ and\ \citenamefont {Druart}(2020)}]{Compere:2020eat}%
  \BibitemOpen
  \bibfield  {author} {\bibinfo {author} {\bibfnamefont {G.}~\bibnamefont {Comp\`ere}}\ and\ \bibinfo {author} {\bibfnamefont {A.}~\bibnamefont {Druart}},\ }\href {\doibase 10.1103/PhysRevD.101.084042} {\bibfield  {journal} {\bibinfo  {journal} {Phys. Rev. D}\ }\textbf {\bibinfo {volume} {101}},\ \bibinfo {pages} {084042} (\bibinfo {year} {2020})},\ \bibinfo {note} {[Erratum: Phys.Rev.D 102, 029901 (2020)]},\ \Eprint {http://arxiv.org/abs/2001.03478} {arXiv:2001.03478 [gr-qc]} \BibitemShut {NoStop}%
\bibitem [{\citenamefont {Kapec}\ and\ \citenamefont {Lupsasca}(2020)}]{Kapec:2019hro}%
  \BibitemOpen
  \bibfield  {author} {\bibinfo {author} {\bibfnamefont {D.}~\bibnamefont {Kapec}}\ and\ \bibinfo {author} {\bibfnamefont {A.}~\bibnamefont {Lupsasca}},\ }\href {\doibase 10.1088/1361-6382/ab519e} {\bibfield  {journal} {\bibinfo  {journal} {Class. Quant. Grav.}\ }\textbf {\bibinfo {volume} {37}},\ \bibinfo {pages} {015006} (\bibinfo {year} {2020})},\ \Eprint {http://arxiv.org/abs/1905.11406} {arXiv:1905.11406 [hep-th]} \BibitemShut {NoStop}%
\bibitem [{\citenamefont {Kinnersley}\ and\ \citenamefont {Walker}(1970)}]{kinnersley1970}%
  \BibitemOpen
  \bibfield  {author} {\bibinfo {author} {\bibfnamefont {W.}~\bibnamefont {Kinnersley}}\ and\ \bibinfo {author} {\bibfnamefont {M.}~\bibnamefont {Walker}},\ }\href {\doibase 10.1103/PhysRevD.2.1359} {\bibfield  {journal} {\bibinfo  {journal} {Phys. Rev. D}\ }\textbf {\bibinfo {volume} {2}},\ \bibinfo {pages} {1359} (\bibinfo {year} {1970})}\BibitemShut {NoStop}%
\bibitem [{\citenamefont {Vilenkin}(1985)}]{vilenkin1985}%
  \BibitemOpen
  \bibfield  {author} {\bibinfo {author} {\bibfnamefont {A.}~\bibnamefont {Vilenkin}},\ }\href {\doibase 10.1016/0370-1573(85)90033-X} {\bibfield  {journal} {\bibinfo  {journal} {Phys. Rept.}\ }\textbf {\bibinfo {volume} {121}},\ \bibinfo {pages} {263} (\bibinfo {year} {1985})}\BibitemShut {NoStop}%
\bibitem [{\citenamefont {Hawking}\ and\ \citenamefont {Ross}(1995)}]{hr}%
  \BibitemOpen
  \bibfield  {author} {\bibinfo {author} {\bibfnamefont {S.~W.}\ \bibnamefont {Hawking}}\ and\ \bibinfo {author} {\bibfnamefont {S.~F.}\ \bibnamefont {Ross}},\ }\href {\doibase 10.1103/PhysRevLett.75.3382} {\bibfield  {journal} {\bibinfo  {journal} {Phys. Rev. Lett.}\ }\textbf {\bibinfo {volume} {75}},\ \bibinfo {pages} {3382} (\bibinfo {year} {1995})},\ \Eprint {http://arxiv.org/abs/gr-qc/9506020} {arXiv:gr-qc/9506020} \BibitemShut {NoStop}%
\bibitem [{\citenamefont {Emparan}(1995)}]{emparan1995}%
  \BibitemOpen
  \bibfield  {author} {\bibinfo {author} {\bibfnamefont {R.}~\bibnamefont {Emparan}},\ }\href {\doibase 10.1103/PhysRevLett.75.3386} {\bibfield  {journal} {\bibinfo  {journal} {Phys. Rev. Lett.}\ }\textbf {\bibinfo {volume} {75}},\ \bibinfo {pages} {3386} (\bibinfo {year} {1995})},\ \Eprint {http://arxiv.org/abs/gr-qc/9506025} {arXiv:gr-qc/9506025} \BibitemShut {NoStop}%
\bibitem [{\citenamefont {Eardley}\ \emph {et~al.}(1995)\citenamefont {Eardley}, \citenamefont {Horowitz}, \citenamefont {Kastor},\ and\ \citenamefont {Traschen}}]{eardley1995}%
  \BibitemOpen
  \bibfield  {author} {\bibinfo {author} {\bibfnamefont {D.~M.}\ \bibnamefont {Eardley}}, \bibinfo {author} {\bibfnamefont {G.~T.}\ \bibnamefont {Horowitz}}, \bibinfo {author} {\bibfnamefont {D.~A.}\ \bibnamefont {Kastor}}, \ and\ \bibinfo {author} {\bibfnamefont {J.~H.}\ \bibnamefont {Traschen}},\ }\href {\doibase 10.1103/PhysRevLett.75.3390} {\bibfield  {journal} {\bibinfo  {journal} {Phys. Rev. Lett.}\ }\textbf {\bibinfo {volume} {75}},\ \bibinfo {pages} {3390} (\bibinfo {year} {1995})},\ \Eprint {http://arxiv.org/abs/gr-qc/9506041} {arXiv:gr-qc/9506041} \BibitemShut {NoStop}%
\bibitem [{\citenamefont {Ashoorioon}\ and\ \citenamefont {Mann}(2014)}]{Ashoorioon:2014ipa}%
  \BibitemOpen
  \bibfield  {author} {\bibinfo {author} {\bibfnamefont {A.}~\bibnamefont {Ashoorioon}}\ and\ \bibinfo {author} {\bibfnamefont {R.~B.}\ \bibnamefont {Mann}},\ }\href {\doibase 10.1088/0264-9381/31/22/225009} {\bibfield  {journal} {\bibinfo  {journal} {Class. Quant. Grav.}\ }\textbf {\bibinfo {volume} {31}},\ \bibinfo {pages} {225009} (\bibinfo {year} {2014})},\ \Eprint {http://arxiv.org/abs/1402.2072} {arXiv:1402.2072 [hep-th]} \BibitemShut {NoStop}%
\bibitem [{\citenamefont {Ashoorioon}\ and\ \citenamefont {Jahani~Poshteh}(2021)}]{ashoorioon2021}%
  \BibitemOpen
  \bibfield  {author} {\bibinfo {author} {\bibfnamefont {A.}~\bibnamefont {Ashoorioon}}\ and\ \bibinfo {author} {\bibfnamefont {M.~B.}\ \bibnamefont {Jahani~Poshteh}},\ }\href {\doibase 10.1016/j.physletb.2021.136224} {\bibfield  {journal} {\bibinfo  {journal} {Phys. Lett. B}\ }\textbf {\bibinfo {volume} {816}},\ \bibinfo {pages} {136224} (\bibinfo {year} {2021})},\ \Eprint {http://arxiv.org/abs/2006.16983} {arXiv:2006.16983 [hep-th]} \BibitemShut {NoStop}%
\bibitem [{\citenamefont {Ashoorioon}\ \emph {et~al.}(2022)\citenamefont {Ashoorioon}, \citenamefont {Jahani~Poshteh},\ and\ \citenamefont {Mann}}]{Ashoorioon:2022zgu}%
  \BibitemOpen
  \bibfield  {author} {\bibinfo {author} {\bibfnamefont {A.}~\bibnamefont {Ashoorioon}}, \bibinfo {author} {\bibfnamefont {M.~B.}\ \bibnamefont {Jahani~Poshteh}}, \ and\ \bibinfo {author} {\bibfnamefont {R.~B.}\ \bibnamefont {Mann}},\ }\href {\doibase 10.1103/PhysRevLett.129.031102} {\bibfield  {journal} {\bibinfo  {journal} {Phys. Rev. Lett.}\ }\textbf {\bibinfo {volume} {129}},\ \bibinfo {pages} {031102} (\bibinfo {year} {2022})},\ \Eprint {http://arxiv.org/abs/2210.10762} {arXiv:2210.10762 [gr-qc]} \BibitemShut {NoStop}%
\bibitem [{\citenamefont {Ashoorioon}\ \emph {et~al.}(2023)\citenamefont {Ashoorioon}, \citenamefont {Jahani~Poshteh},\ and\ \citenamefont {Mann}}]{Ashoorioon:2021gjs}%
  \BibitemOpen
  \bibfield  {author} {\bibinfo {author} {\bibfnamefont {A.}~\bibnamefont {Ashoorioon}}, \bibinfo {author} {\bibfnamefont {M.~B.}\ \bibnamefont {Jahani~Poshteh}}, \ and\ \bibinfo {author} {\bibfnamefont {R.~B.}\ \bibnamefont {Mann}},\ }\href {\doibase 10.1103/PhysRevD.107.044031} {\bibfield  {journal} {\bibinfo  {journal} {Phys. Rev. D}\ }\textbf {\bibinfo {volume} {107}},\ \bibinfo {pages} {044031} (\bibinfo {year} {2023})},\ \Eprint {http://arxiv.org/abs/2110.13132} {arXiv:2110.13132 [gr-qc]} \BibitemShut {NoStop}%
\bibitem [{\citenamefont {Narayan}\ and\ \citenamefont {Popham}(1993)}]{Narayan:1993bd}%
  \BibitemOpen
  \bibfield  {author} {\bibinfo {author} {\bibfnamefont {R.}~\bibnamefont {Narayan}}\ and\ \bibinfo {author} {\bibfnamefont {R.}~\bibnamefont {Popham}},\ }\href {\doibase 10.1038/362820a0} {\bibfield  {journal} {\bibinfo  {journal} {Nature}\ }\textbf {\bibinfo {volume} {362}},\ \bibinfo {pages} {820} (\bibinfo {year} {1993})}\BibitemShut {NoStop}%
\bibitem [{\citenamefont {Tanaka}\ \emph {et~al.}(1995)\citenamefont {Tanaka} \emph {et~al.}}]{Tanaka:1995en}%
  \BibitemOpen
  \bibfield  {author} {\bibinfo {author} {\bibfnamefont {Y.}~\bibnamefont {Tanaka}} \emph {et~al.},\ }\href {\doibase 10.1038/375659a0} {\bibfield  {journal} {\bibinfo  {journal} {Nature}\ }\textbf {\bibinfo {volume} {375}},\ \bibinfo {pages} {659} (\bibinfo {year} {1995})}\BibitemShut {NoStop}%
\bibitem [{\citenamefont {Kato}\ \emph {et~al.}(2022)\citenamefont {Kato}, \citenamefont {Ebisuzaki},\ and\ \citenamefont {Tajima}}]{Kato:2022cur}%
  \BibitemOpen
  \bibfield  {author} {\bibinfo {author} {\bibfnamefont {Y.}~\bibnamefont {Kato}}, \bibinfo {author} {\bibfnamefont {T.}~\bibnamefont {Ebisuzaki}}, \ and\ \bibinfo {author} {\bibfnamefont {T.}~\bibnamefont {Tajima}},\ }\href {\doibase 10.3847/1538-4357/ac56e3} {\bibfield  {journal} {\bibinfo  {journal} {Astrophys. J.}\ }\textbf {\bibinfo {volume} {929}},\ \bibinfo {pages} {42} (\bibinfo {year} {2022})},\ \Eprint {http://arxiv.org/abs/2201.11755} {arXiv:2201.11755 [astro-ph.HE]} \BibitemShut {NoStop}%
\bibitem [{\citenamefont {Kadowaki}\ \emph {et~al.}(2018)\citenamefont {Kadowaki}, \citenamefont {de~Gouveia Dal~Pino},\ and\ \citenamefont {Stone}}]{Kadowaki:2018dtb}%
  \BibitemOpen
  \bibfield  {author} {\bibinfo {author} {\bibfnamefont {L.~H.~S.}\ \bibnamefont {Kadowaki}}, \bibinfo {author} {\bibfnamefont {E.~M.}\ \bibnamefont {de~Gouveia Dal~Pino}}, \ and\ \bibinfo {author} {\bibfnamefont {J.~M.}\ \bibnamefont {Stone}},\ }\href {\doibase 10.3847/1538-4357/aad4ff} {\bibfield  {journal} {\bibinfo  {journal} {Astrophys. J.}\ }\textbf {\bibinfo {volume} {864}},\ \bibinfo {pages} {52} (\bibinfo {year} {2018})},\ \Eprint {http://arxiv.org/abs/1803.08557} {arXiv:1803.08557 [astro-ph.HE]} \BibitemShut {NoStop}%
\bibitem [{\citenamefont {Andreoni}\ \emph {et~al.}(2022)\citenamefont {Andreoni} \emph {et~al.}}]{Andreoni:2022afu}%
  \BibitemOpen
  \bibfield  {author} {\bibinfo {author} {\bibfnamefont {I.}~\bibnamefont {Andreoni}} \emph {et~al.},\ }\href {\doibase 10.1038/s41586-022-05465-8} {\bibfield  {journal} {\bibinfo  {journal} {Nature}\ }\textbf {\bibinfo {volume} {612}},\ \bibinfo {pages} {430} (\bibinfo {year} {2022})},\ \bibinfo {note} {[Erratum: Nature 613, E6 (2023)]},\ \Eprint {http://arxiv.org/abs/2211.16530} {arXiv:2211.16530 [astro-ph.HE]} \BibitemShut {NoStop}%
\bibitem [{\citenamefont {Kosec}\ \emph {et~al.}(2023)\citenamefont {Kosec} \emph {et~al.}}]{Kosec:2023qva}%
  \BibitemOpen
  \bibfield  {author} {\bibinfo {author} {\bibfnamefont {P.}~\bibnamefont {Kosec}} \emph {et~al.},\ }\href {\doibase 10.1038/s41550-023-01929-7} {\bibfield  {journal} {\bibinfo  {journal} {Nature Astron.}\ }\textbf {\bibinfo {volume} {7}},\ \bibinfo {pages} {715} (\bibinfo {year} {2023})},\ \Eprint {http://arxiv.org/abs/2304.05490} {arXiv:2304.05490 [astro-ph.HE]} \BibitemShut {NoStop}%
\bibitem [{\citenamefont {Abramowicz}\ and\ \citenamefont {Fragile}(2013)}]{Abramowicz:2011xu}%
  \BibitemOpen
  \bibfield  {author} {\bibinfo {author} {\bibfnamefont {M.~A.}\ \bibnamefont {Abramowicz}}\ and\ \bibinfo {author} {\bibfnamefont {P.~C.}\ \bibnamefont {Fragile}},\ }\href {\doibase 10.12942/lrr-2013-1} {\bibfield  {journal} {\bibinfo  {journal} {Living Rev. Rel.}\ }\textbf {\bibinfo {volume} {16}},\ \bibinfo {pages} {1} (\bibinfo {year} {2013})},\ \Eprint {http://arxiv.org/abs/1104.5499} {arXiv:1104.5499 [astro-ph.HE]} \BibitemShut {NoStop}%
\bibitem [{\citenamefont {Joshi}\ \emph {et~al.}(2014)\citenamefont {Joshi}, \citenamefont {Malafarina},\ and\ \citenamefont {Narayan}}]{Joshi:2013dva}%
  \BibitemOpen
  \bibfield  {author} {\bibinfo {author} {\bibfnamefont {P.~S.}\ \bibnamefont {Joshi}}, \bibinfo {author} {\bibfnamefont {D.}~\bibnamefont {Malafarina}}, \ and\ \bibinfo {author} {\bibfnamefont {R.}~\bibnamefont {Narayan}},\ }\href {\doibase 10.1088/0264-9381/31/1/015002} {\bibfield  {journal} {\bibinfo  {journal} {Class. Quant. Grav.}\ }\textbf {\bibinfo {volume} {31}},\ \bibinfo {pages} {015002} (\bibinfo {year} {2014})},\ \Eprint {http://arxiv.org/abs/1304.7331} {arXiv:1304.7331 [gr-qc]} \BibitemShut {NoStop}%
\bibitem [{\citenamefont {Shaikh}\ and\ \citenamefont {Joshi}(2019)}]{Shaikh:2019hbm}%
  \BibitemOpen
  \bibfield  {author} {\bibinfo {author} {\bibfnamefont {R.}~\bibnamefont {Shaikh}}\ and\ \bibinfo {author} {\bibfnamefont {P.~S.}\ \bibnamefont {Joshi}},\ }\href {\doibase 10.1088/1475-7516/2019/10/064} {\bibfield  {journal} {\bibinfo  {journal} {JCAP}\ }\textbf {\bibinfo {volume} {10}},\ \bibinfo {pages} {064} (\bibinfo {year} {2019})},\ \Eprint {http://arxiv.org/abs/1909.10322} {arXiv:1909.10322 [gr-qc]} \BibitemShut {NoStop}%
\bibitem [{\citenamefont {Stuchlik}(2005)}]{Stuchlik:2005euw}%
  \BibitemOpen
  \bibfield  {author} {\bibinfo {author} {\bibfnamefont {Z.}~\bibnamefont {Stuchlik}},\ }\href {\doibase 10.1142/S0217732305016865} {\bibfield  {journal} {\bibinfo  {journal} {Mod. Phys. Lett. A}\ }\textbf {\bibinfo {volume} {20}},\ \bibinfo {pages} {561} (\bibinfo {year} {2005})},\ \Eprint {http://arxiv.org/abs/0804.2266} {arXiv:0804.2266 [astro-ph]} \BibitemShut {NoStop}%
\bibitem [{\citenamefont {Boshkayev}\ \emph {et~al.}(2020)\citenamefont {Boshkayev}, \citenamefont {Idrissov}, \citenamefont {Luongo},\ and\ \citenamefont {Malafarina}}]{Boshkayev:2020kle}%
  \BibitemOpen
  \bibfield  {author} {\bibinfo {author} {\bibfnamefont {K.}~\bibnamefont {Boshkayev}}, \bibinfo {author} {\bibfnamefont {A.}~\bibnamefont {Idrissov}}, \bibinfo {author} {\bibfnamefont {O.}~\bibnamefont {Luongo}}, \ and\ \bibinfo {author} {\bibfnamefont {D.}~\bibnamefont {Malafarina}},\ }\href {\doibase 10.1093/mnras/staa1564} {\bibfield  {journal} {\bibinfo  {journal} {Mon. Not. Roy. Astron. Soc.}\ }\textbf {\bibinfo {volume} {496}},\ \bibinfo {pages} {1115} (\bibinfo {year} {2020})},\ \Eprint {http://arxiv.org/abs/2006.01269} {arXiv:2006.01269 [astro-ph.HE]} \BibitemShut {NoStop}%
\bibitem [{\citenamefont {Kurmanov}\ \emph {et~al.}(2022)\citenamefont {Kurmanov}, \citenamefont {Boshkayev}, \citenamefont {Giamb\`o}, \citenamefont {Konysbayev}, \citenamefont {Luongo}, \citenamefont {Malafarina},\ and\ \citenamefont {Quevedo}}]{Kurmanov:2021uqv}%
  \BibitemOpen
  \bibfield  {author} {\bibinfo {author} {\bibfnamefont {E.}~\bibnamefont {Kurmanov}}, \bibinfo {author} {\bibfnamefont {K.}~\bibnamefont {Boshkayev}}, \bibinfo {author} {\bibfnamefont {R.}~\bibnamefont {Giamb\`o}}, \bibinfo {author} {\bibfnamefont {T.}~\bibnamefont {Konysbayev}}, \bibinfo {author} {\bibfnamefont {O.}~\bibnamefont {Luongo}}, \bibinfo {author} {\bibfnamefont {D.}~\bibnamefont {Malafarina}}, \ and\ \bibinfo {author} {\bibfnamefont {H.}~\bibnamefont {Quevedo}},\ }\href {\doibase 10.3847/1538-4357/ac41d4} {\bibfield  {journal} {\bibinfo  {journal} {Astrophys. J.}\ }\textbf {\bibinfo {volume} {925}},\ \bibinfo {pages} {210} (\bibinfo {year} {2022})},\ \Eprint {http://arxiv.org/abs/2110.15402} {arXiv:2110.15402 [astro-ph.HE]} \BibitemShut {NoStop}%
\bibitem [{\citenamefont {Boshkayev}\ \emph {et~al.}(2022)\citenamefont {Boshkayev}, \citenamefont {Konysbayev}, \citenamefont {Kurmanov}, \citenamefont {Luongo},\ and\ \citenamefont {Malafarina}}]{Boshkayev:2022vlv}%
  \BibitemOpen
  \bibfield  {author} {\bibinfo {author} {\bibfnamefont {K.}~\bibnamefont {Boshkayev}}, \bibinfo {author} {\bibfnamefont {T.}~\bibnamefont {Konysbayev}}, \bibinfo {author} {\bibfnamefont {Y.}~\bibnamefont {Kurmanov}}, \bibinfo {author} {\bibfnamefont {O.}~\bibnamefont {Luongo}}, \ and\ \bibinfo {author} {\bibfnamefont {D.}~\bibnamefont {Malafarina}},\ }\href {\doibase 10.3847/1538-4357/ac8804} {\bibfield  {journal} {\bibinfo  {journal} {Astrophys. J.}\ }\textbf {\bibinfo {volume} {936}},\ \bibinfo {pages} {96} (\bibinfo {year} {2022})},\ \Eprint {http://arxiv.org/abs/2205.04208} {arXiv:2205.04208 [gr-qc]} \BibitemShut {NoStop}%
\bibitem [{\citenamefont {D'Agostino}\ \emph {et~al.}(2023)\citenamefont {D'Agostino}, \citenamefont {Giamb\`o},\ and\ \citenamefont {Luongo}}]{DAgostino:2022ckg}%
  \BibitemOpen
  \bibfield  {author} {\bibinfo {author} {\bibfnamefont {R.}~\bibnamefont {D'Agostino}}, \bibinfo {author} {\bibfnamefont {R.}~\bibnamefont {Giamb\`o}}, \ and\ \bibinfo {author} {\bibfnamefont {O.}~\bibnamefont {Luongo}},\ }\href {\doibase 10.1103/PhysRevD.107.043032} {\bibfield  {journal} {\bibinfo  {journal} {Phys. Rev. D}\ }\textbf {\bibinfo {volume} {107}},\ \bibinfo {pages} {043032} (\bibinfo {year} {2023})},\ \Eprint {http://arxiv.org/abs/2204.02098} {arXiv:2204.02098 [gr-qc]} \BibitemShut {NoStop}%
\bibitem [{\citenamefont {Boshkayev}\ \emph {et~al.}(2024{\natexlab{a}})\citenamefont {Boshkayev}, \citenamefont {Konysbayev}, \citenamefont {Kurmanov}, \citenamefont {Luongo}, \citenamefont {Muccino}, \citenamefont {Taukenova},\ and\ \citenamefont {Urazalina}}]{Boshkayev:2023fft}%
  \BibitemOpen
  \bibfield  {author} {\bibinfo {author} {\bibfnamefont {K.}~\bibnamefont {Boshkayev}}, \bibinfo {author} {\bibfnamefont {T.}~\bibnamefont {Konysbayev}}, \bibinfo {author} {\bibfnamefont {Y.}~\bibnamefont {Kurmanov}}, \bibinfo {author} {\bibfnamefont {O.}~\bibnamefont {Luongo}}, \bibinfo {author} {\bibfnamefont {M.}~\bibnamefont {Muccino}}, \bibinfo {author} {\bibfnamefont {A.}~\bibnamefont {Taukenova}}, \ and\ \bibinfo {author} {\bibfnamefont {A.}~\bibnamefont {Urazalina}},\ }\href {\doibase 10.1140/epjc/s10052-024-12446-w} {\bibfield  {journal} {\bibinfo  {journal} {Eur. Phys. J. C}\ }\textbf {\bibinfo {volume} {84}},\ \bibinfo {pages} {230} (\bibinfo {year} {2024}{\natexlab{a}})},\ \Eprint {http://arxiv.org/abs/2307.15003} {arXiv:2307.15003 [gr-qc]} \BibitemShut {NoStop}%
\bibitem [{\citenamefont {Boshkayev}\ \emph {et~al.}(2024{\natexlab{b}})\citenamefont {Boshkayev}, \citenamefont {Konysbayev}, \citenamefont {Kurmanov}, \citenamefont {Luongo}, \citenamefont {Muccino}, \citenamefont {Quevedo},\ and\ \citenamefont {Urazalina}}]{Boshkayev:2023ipb}%
  \BibitemOpen
  \bibfield  {author} {\bibinfo {author} {\bibfnamefont {K.}~\bibnamefont {Boshkayev}}, \bibinfo {author} {\bibfnamefont {T.}~\bibnamefont {Konysbayev}}, \bibinfo {author} {\bibfnamefont {Y.}~\bibnamefont {Kurmanov}}, \bibinfo {author} {\bibfnamefont {O.}~\bibnamefont {Luongo}}, \bibinfo {author} {\bibfnamefont {M.}~\bibnamefont {Muccino}}, \bibinfo {author} {\bibfnamefont {H.}~\bibnamefont {Quevedo}}, \ and\ \bibinfo {author} {\bibfnamefont {A.}~\bibnamefont {Urazalina}},\ }\href {\doibase 10.1140/epjp/s13360-024-05072-8} {\bibfield  {journal} {\bibinfo  {journal} {Eur. Phys. J. Plus}\ }\textbf {\bibinfo {volume} {139}},\ \bibinfo {pages} {273} (\bibinfo {year} {2024}{\natexlab{b}})},\ \Eprint {http://arxiv.org/abs/2306.15050} {arXiv:2306.15050 [gr-qc]} \BibitemShut {NoStop}%
\bibitem [{\citenamefont {Boshkayev}\ \emph {et~al.}(2021)\citenamefont {Boshkayev}, \citenamefont {Konysbayev}, \citenamefont {Kurmanov}, \citenamefont {Luongo}, \citenamefont {Malafarina},\ and\ \citenamefont {Quevedo}}]{Boshkayev:2021chc}%
  \BibitemOpen
  \bibfield  {author} {\bibinfo {author} {\bibfnamefont {K.}~\bibnamefont {Boshkayev}}, \bibinfo {author} {\bibfnamefont {T.}~\bibnamefont {Konysbayev}}, \bibinfo {author} {\bibfnamefont {E.}~\bibnamefont {Kurmanov}}, \bibinfo {author} {\bibfnamefont {O.}~\bibnamefont {Luongo}}, \bibinfo {author} {\bibfnamefont {D.}~\bibnamefont {Malafarina}}, \ and\ \bibinfo {author} {\bibfnamefont {H.}~\bibnamefont {Quevedo}},\ }\href {\doibase 10.1103/PhysRevD.104.084009} {\bibfield  {journal} {\bibinfo  {journal} {Phys. Rev. D}\ }\textbf {\bibinfo {volume} {104}},\ \bibinfo {pages} {084009} (\bibinfo {year} {2021})},\ \Eprint {http://arxiv.org/abs/2106.04932} {arXiv:2106.04932 [gr-qc]} \BibitemShut {NoStop}%
\bibitem [{\citenamefont {Jahani~Poshteh}(2022)}]{JahaniPoshteh:2022yei}%
  \BibitemOpen
  \bibfield  {author} {\bibinfo {author} {\bibfnamefont {M.~B.}\ \bibnamefont {Jahani~Poshteh}},\ }\href {\doibase 10.1103/PhysRevD.106.044037} {\bibfield  {journal} {\bibinfo  {journal} {Phys. Rev. D}\ }\textbf {\bibinfo {volume} {106}},\ \bibinfo {pages} {044037} (\bibinfo {year} {2022})},\ \bibinfo {note} {[Erratum: Phys.Rev.D 107, 129901 (2023)]},\ \Eprint {http://arxiv.org/abs/2203.03223} {arXiv:2203.03223 [gr-qc]} \BibitemShut {NoStop}%
\bibitem [{\citenamefont {Morris}\ \emph {et~al.}(2017)\citenamefont {Morris}, \citenamefont {Zhao},\ and\ \citenamefont {Goss}}]{morris2017}%
  \BibitemOpen
  \bibfield  {author} {\bibinfo {author} {\bibfnamefont {M.~R.}\ \bibnamefont {Morris}}, \bibinfo {author} {\bibfnamefont {J.-H.}\ \bibnamefont {Zhao}}, \ and\ \bibinfo {author} {\bibfnamefont {W.}~\bibnamefont {Goss}},\ }\href {https://iopscience.iop.org/article/10.3847/2041-8213/aa9985/meta} {\bibfield  {journal} {\bibinfo  {journal} {The Astrophysical Journal Letters}\ }\textbf {\bibinfo {volume} {850}},\ \bibinfo {pages} {L23} (\bibinfo {year} {2017})}\BibitemShut {NoStop}%
\bibitem [{\citenamefont {Vilenkin}\ \emph {et~al.}(2018)\citenamefont {Vilenkin}, \citenamefont {Levin},\ and\ \citenamefont {Gruzinov}}]{vilenkin2018}%
  \BibitemOpen
  \bibfield  {author} {\bibinfo {author} {\bibfnamefont {A.}~\bibnamefont {Vilenkin}}, \bibinfo {author} {\bibfnamefont {Y.}~\bibnamefont {Levin}}, \ and\ \bibinfo {author} {\bibfnamefont {A.}~\bibnamefont {Gruzinov}},\ }\href {\doibase 10.1088/1475-7516/2018/11/008} {\bibfield  {journal} {\bibinfo  {journal} {JCAP}\ }\textbf {\bibinfo {volume} {11}},\ \bibinfo {pages} {008} (\bibinfo {year} {2018})},\ \Eprint {http://arxiv.org/abs/1808.00670} {arXiv:1808.00670 [astro-ph.CO]} \BibitemShut {NoStop}%
\bibitem [{\citenamefont {Novikov}\ and\ \citenamefont {Thorne}(1973)}]{Novikov:1973}%
  \BibitemOpen
  \bibfield  {author} {\bibinfo {author} {\bibfnamefont {I.~D.}\ \bibnamefont {Novikov}}\ and\ \bibinfo {author} {\bibfnamefont {K.~S.}\ \bibnamefont {Thorne}},\ }in\ \href@noop {} {\emph {\bibinfo {booktitle} {Black Holes}}},\ \bibinfo {editor} {edited by\ \bibinfo {editor} {\bibfnamefont {C.}~\bibnamefont {DeWitt}}\ and\ \bibinfo {editor} {\bibfnamefont {B.}~\bibnamefont {DeWitt}}}\ (\bibinfo  {publisher} {Gordon and Breach},\ \bibinfo {address} {New York},\ \bibinfo {year} {1973})\ pp.\ \bibinfo {pages} {343--450}\BibitemShut {NoStop}%
\bibitem [{\citenamefont {Page}\ and\ \citenamefont {Thorne}(1974)}]{Page:1974he}%
  \BibitemOpen
  \bibfield  {author} {\bibinfo {author} {\bibfnamefont {D.~N.}\ \bibnamefont {Page}}\ and\ \bibinfo {author} {\bibfnamefont {K.~S.}\ \bibnamefont {Thorne}},\ }\href {\doibase 10.1086/152990} {\bibfield  {journal} {\bibinfo  {journal} {Astrophys. J.}\ }\textbf {\bibinfo {volume} {191}},\ \bibinfo {pages} {499} (\bibinfo {year} {1974})}\BibitemShut {NoStop}%
\bibitem [{\citenamefont {Thorne}(1974)}]{Thorne:1974ve}%
  \BibitemOpen
  \bibfield  {author} {\bibinfo {author} {\bibfnamefont {K.~S.}\ \bibnamefont {Thorne}},\ }\href {\doibase 10.1086/152991} {\bibfield  {journal} {\bibinfo  {journal} {Astrophys. J.}\ }\textbf {\bibinfo {volume} {191}},\ \bibinfo {pages} {507} (\bibinfo {year} {1974})}\BibitemShut {NoStop}%
\bibitem [{\citenamefont {Pleba{\'n}ski}\ and\ \citenamefont {Demia{\'n}ski}(1976)}]{plebanski1976}%
  \BibitemOpen
  \bibfield  {author} {\bibinfo {author} {\bibfnamefont {J.~F.}\ \bibnamefont {Pleba{\'n}ski}}\ and\ \bibinfo {author} {\bibfnamefont {M.}~\bibnamefont {Demia{\'n}ski}},\ }\href {\doibase 10.1016/0003-4916(76)90240-2} {\bibfield  {journal} {\bibinfo  {journal} {Annals Phys.}\ }\textbf {\bibinfo {volume} {98}},\ \bibinfo {pages} {98} (\bibinfo {year} {1976})}\BibitemShut {NoStop}%
\bibitem [{\citenamefont {Van~den Bergh}\ and\ \citenamefont {Carminati}(2020)}]{VandenBergh:2020lvf}%
  \BibitemOpen
  \bibfield  {author} {\bibinfo {author} {\bibfnamefont {N.}~\bibnamefont {Van~den Bergh}}\ and\ \bibinfo {author} {\bibfnamefont {J.}~\bibnamefont {Carminati}},\ }\href {\doibase 10.1088/1361-6382/abbba3} {\bibfield  {journal} {\bibinfo  {journal} {Class. Quant. Grav.}\ }\textbf {\bibinfo {volume} {37}},\ \bibinfo {pages} {215010} (\bibinfo {year} {2020})},\ \Eprint {http://arxiv.org/abs/2009.11516} {arXiv:2009.11516 [gr-qc]} \BibitemShut {NoStop}%
\bibitem [{\citenamefont {Griffiths}\ and\ \citenamefont {Podolsky}(2006)}]{Griffiths:2005qp}%
  \BibitemOpen
  \bibfield  {author} {\bibinfo {author} {\bibfnamefont {J.~B.}\ \bibnamefont {Griffiths}}\ and\ \bibinfo {author} {\bibfnamefont {J.}~\bibnamefont {Podolsky}},\ }\href {\doibase 10.1142/S0218271806007742} {\bibfield  {journal} {\bibinfo  {journal} {Int. J. Mod. Phys. D}\ }\textbf {\bibinfo {volume} {15}},\ \bibinfo {pages} {335} (\bibinfo {year} {2006})},\ \Eprint {http://arxiv.org/abs/gr-qc/0511091} {arXiv:gr-qc/0511091} \BibitemShut {NoStop}%
\bibitem [{\citenamefont {Podolsky}\ and\ \citenamefont {Griffiths}(2006)}]{Podolsky:2006px}%
  \BibitemOpen
  \bibfield  {author} {\bibinfo {author} {\bibfnamefont {J.}~\bibnamefont {Podolsky}}\ and\ \bibinfo {author} {\bibfnamefont {J.~B.}\ \bibnamefont {Griffiths}},\ }\href {\doibase 10.1103/PhysRevD.73.044018} {\bibfield  {journal} {\bibinfo  {journal} {Phys. Rev. D}\ }\textbf {\bibinfo {volume} {73}},\ \bibinfo {pages} {044018} (\bibinfo {year} {2006})},\ \Eprint {http://arxiv.org/abs/gr-qc/0601130} {arXiv:gr-qc/0601130} \BibitemShut {NoStop}%
\bibitem [{\citenamefont {Lim}(2014)}]{lim2014}%
  \BibitemOpen
  \bibfield  {author} {\bibinfo {author} {\bibfnamefont {Y.-K.}\ \bibnamefont {Lim}},\ }\href {\doibase 10.1103/PhysRevD.89.104016} {\bibfield  {journal} {\bibinfo  {journal} {Phys. Rev. D}\ }\textbf {\bibinfo {volume} {89}},\ \bibinfo {pages} {104016} (\bibinfo {year} {2014})},\ \Eprint {http://arxiv.org/abs/1405.2611} {arXiv:1405.2611 [gr-qc]} \BibitemShut {NoStop}%
\bibitem [{\citenamefont {Howes}(1979)}]{howes1979existence}%
  \BibitemOpen
  \bibfield  {author} {\bibinfo {author} {\bibfnamefont {R.}~\bibnamefont {Howes}},\ }\href {https://www.publish.csiro.au/ph/PH790293} {\bibfield  {journal} {\bibinfo  {journal} {Australian Journal of Physics}\ }\textbf {\bibinfo {volume} {32}},\ \bibinfo {pages} {293} (\bibinfo {year} {1979})}\BibitemShut {NoStop}%
\bibitem [{\citenamefont {Boonserm}\ \emph {et~al.}(2020)\citenamefont {Boonserm}, \citenamefont {Ngampitipan}, \citenamefont {Simpson},\ and\ \citenamefont {Visser}}]{Boonserm:2019nqq}%
  \BibitemOpen
  \bibfield  {author} {\bibinfo {author} {\bibfnamefont {P.}~\bibnamefont {Boonserm}}, \bibinfo {author} {\bibfnamefont {T.}~\bibnamefont {Ngampitipan}}, \bibinfo {author} {\bibfnamefont {A.}~\bibnamefont {Simpson}}, \ and\ \bibinfo {author} {\bibfnamefont {M.}~\bibnamefont {Visser}},\ }\href {\doibase 10.1103/PhysRevD.101.024050} {\bibfield  {journal} {\bibinfo  {journal} {Phys. Rev. D}\ }\textbf {\bibinfo {volume} {101}},\ \bibinfo {pages} {024050} (\bibinfo {year} {2020})},\ \Eprint {http://arxiv.org/abs/1909.06755} {arXiv:1909.06755 [gr-qc]} \BibitemShut {NoStop}%
\bibitem [{\citenamefont {Padmanabhan}(2003)}]{Padmanabhan:2002ji}%
  \BibitemOpen
  \bibfield  {author} {\bibinfo {author} {\bibfnamefont {T.}~\bibnamefont {Padmanabhan}},\ }\href {\doibase 10.1016/S0370-1573(03)00120-0} {\bibfield  {journal} {\bibinfo  {journal} {Phys. Rept.}\ }\textbf {\bibinfo {volume} {380}},\ \bibinfo {pages} {235} (\bibinfo {year} {2003})},\ \Eprint {http://arxiv.org/abs/hep-th/0212290} {arXiv:hep-th/0212290} \BibitemShut {NoStop}%
\bibitem [{\citenamefont {Bambi}\ and\ \citenamefont {Nampalliwar}(2016)}]{Bambi:2016iip}%
  \BibitemOpen
  \bibfield  {author} {\bibinfo {author} {\bibfnamefont {C.}~\bibnamefont {Bambi}}\ and\ \bibinfo {author} {\bibfnamefont {S.}~\bibnamefont {Nampalliwar}},\ }\href {\doibase 10.1209/0295-5075/116/30006} {\bibfield  {journal} {\bibinfo  {journal} {EPL}\ }\textbf {\bibinfo {volume} {116}},\ \bibinfo {pages} {30006} (\bibinfo {year} {2016})},\ \Eprint {http://arxiv.org/abs/1604.02643} {arXiv:1604.02643 [gr-qc]} \BibitemShut {NoStop}%
\bibitem [{\citenamefont {Eisenhauer}\ \emph {et~al.}(2003)\citenamefont {Eisenhauer}, \citenamefont {Sch{\"o}del}, \citenamefont {Genzel}, \citenamefont {Ott}, \citenamefont {Tecza}, \citenamefont {Abuter}, \citenamefont {Eckart},\ and\ \citenamefont {Alexander}}]{eisenhauer2003geometric}%
  \BibitemOpen
  \bibfield  {author} {\bibinfo {author} {\bibfnamefont {F.}~\bibnamefont {Eisenhauer}}, \bibinfo {author} {\bibfnamefont {R.}~\bibnamefont {Sch{\"o}del}}, \bibinfo {author} {\bibfnamefont {R.}~\bibnamefont {Genzel}}, \bibinfo {author} {\bibfnamefont {T.}~\bibnamefont {Ott}}, \bibinfo {author} {\bibfnamefont {M.}~\bibnamefont {Tecza}}, \bibinfo {author} {\bibfnamefont {R.}~\bibnamefont {Abuter}}, \bibinfo {author} {\bibfnamefont {A.}~\bibnamefont {Eckart}}, \ and\ \bibinfo {author} {\bibfnamefont {T.}~\bibnamefont {Alexander}},\ }\href {https://iopscience.iop.org/article/10.1086/380188/meta} {\bibfield  {journal} {\bibinfo  {journal} {The Astrophysical Journal Letters}\ }\textbf {\bibinfo {volume} {597}},\ \bibinfo {pages} {L121} (\bibinfo {year} {2003})}\BibitemShut {NoStop}%
\bibitem [{\citenamefont {Bardeen}\ and\ \citenamefont {Petterson}(1975)}]{Bardeen:1975zz}%
  \BibitemOpen
  \bibfield  {author} {\bibinfo {author} {\bibfnamefont {J.~M.}\ \bibnamefont {Bardeen}}\ and\ \bibinfo {author} {\bibfnamefont {J.~A.}\ \bibnamefont {Petterson}},\ }\href {\doibase 10.1086/181711} {\bibfield  {journal} {\bibinfo  {journal} {Astrophys. J. Lett.}\ }\textbf {\bibinfo {volume} {195}},\ \bibinfo {pages} {L65} (\bibinfo {year} {1975})}\BibitemShut {NoStop}%
\bibitem [{\citenamefont {Mummery}\ and\ \citenamefont {Balbus}(2023)}]{Mummery:2023tgh}%
  \BibitemOpen
  \bibfield  {author} {\bibinfo {author} {\bibfnamefont {A.}~\bibnamefont {Mummery}}\ and\ \bibinfo {author} {\bibfnamefont {S.}~\bibnamefont {Balbus}},\ }\href {\doibase 10.1093/mnras/stad641} {\bibfield  {journal} {\bibinfo  {journal} {Mon. Not. Roy. Astron. Soc.}\ }\textbf {\bibinfo {volume} {521}},\ \bibinfo {pages} {2439} (\bibinfo {year} {2023})},\ \Eprint {http://arxiv.org/abs/2302.14437} {arXiv:2302.14437 [astro-ph.HE]} \BibitemShut {NoStop}%
\bibitem [{\citenamefont {Stuchlik}\ and\ \citenamefont {Hledik}(1999)}]{Stuchlik:1999qk}%
  \BibitemOpen
  \bibfield  {author} {\bibinfo {author} {\bibfnamefont {Z.}~\bibnamefont {Stuchlik}}\ and\ \bibinfo {author} {\bibfnamefont {S.}~\bibnamefont {Hledik}},\ }\href {\doibase 10.1103/PhysRevD.60.044006} {\bibfield  {journal} {\bibinfo  {journal} {Phys. Rev. D}\ }\textbf {\bibinfo {volume} {60}},\ \bibinfo {pages} {044006} (\bibinfo {year} {1999})}\BibitemShut {NoStop}%
\bibitem [{\citenamefont {Li}\ \emph {et~al.}(2005)\citenamefont {Li}, \citenamefont {Zimmerman}, \citenamefont {Narayan},\ and\ \citenamefont {McClintock}}]{Li:2004aq}%
  \BibitemOpen
  \bibfield  {author} {\bibinfo {author} {\bibfnamefont {L.-X.}\ \bibnamefont {Li}}, \bibinfo {author} {\bibfnamefont {E.~R.}\ \bibnamefont {Zimmerman}}, \bibinfo {author} {\bibfnamefont {R.}~\bibnamefont {Narayan}}, \ and\ \bibinfo {author} {\bibfnamefont {J.~E.}\ \bibnamefont {McClintock}},\ }\href {\doibase 10.1086/428089} {\bibfield  {journal} {\bibinfo  {journal} {Astrophys. J. Suppl.}\ }\textbf {\bibinfo {volume} {157}},\ \bibinfo {pages} {335} (\bibinfo {year} {2005})},\ \Eprint {http://arxiv.org/abs/astro-ph/0411583} {arXiv:astro-ph/0411583} \BibitemShut {NoStop}%
\bibitem [{\citenamefont {Reynolds}(2014)}]{Reynolds:2013qqa}%
  \BibitemOpen
  \bibfield  {author} {\bibinfo {author} {\bibfnamefont {C.~S.}\ \bibnamefont {Reynolds}},\ }\href {\doibase 10.1007/s11214-013-0006-6} {\bibfield  {journal} {\bibinfo  {journal} {Space Sci. Rev.}\ }\textbf {\bibinfo {volume} {183}},\ \bibinfo {pages} {277} (\bibinfo {year} {2014})},\ \Eprint {http://arxiv.org/abs/1302.3260} {arXiv:1302.3260 [astro-ph.HE]} \BibitemShut {NoStop}%
\bibitem [{\citenamefont {Narzilloev}\ and\ \citenamefont {Ahmedov}(2022)}]{Narzilloev:2022avv}%
  \BibitemOpen
  \bibfield  {author} {\bibinfo {author} {\bibfnamefont {B.}~\bibnamefont {Narzilloev}}\ and\ \bibinfo {author} {\bibfnamefont {B.}~\bibnamefont {Ahmedov}},\ }\href {\doibase 10.3390/sym14091765} {\bibfield  {journal} {\bibinfo  {journal} {Symmetry}\ }\textbf {\bibinfo {volume} {14}},\ \bibinfo {pages} {1765} (\bibinfo {year} {2022})}\BibitemShut {NoStop}%
\bibitem [{\citenamefont {Poisson}\ \emph {et~al.}(2011)\citenamefont {Poisson}, \citenamefont {Pound},\ and\ \citenamefont {Vega}}]{Poisson:2011nh}%
  \BibitemOpen
  \bibfield  {author} {\bibinfo {author} {\bibfnamefont {E.}~\bibnamefont {Poisson}}, \bibinfo {author} {\bibfnamefont {A.}~\bibnamefont {Pound}}, \ and\ \bibinfo {author} {\bibfnamefont {I.}~\bibnamefont {Vega}},\ }\href {\doibase 10.12942/lrr-2011-7} {\bibfield  {journal} {\bibinfo  {journal} {Living Rev. Rel.}\ }\textbf {\bibinfo {volume} {14}},\ \bibinfo {pages} {7} (\bibinfo {year} {2011})},\ \Eprint {http://arxiv.org/abs/1102.0529} {arXiv:1102.0529 [gr-qc]} \BibitemShut {NoStop}%
\end{thebibliography}
\end{document}